\documentclass[%
 reprint,
showpacs,
 amsmath,amssymb,
 aps,
]{revtex4-1}

\usepackage{graphicx}
\usepackage{dcolumn}
\usepackage{bm}
\usepackage{color}
\usepackage{enumitem}
\usepackage{textcomp} 

\usepackage[
margin=0.65in,top=0.1in,
]{geometry}

\begin{document}

\preprint{APS/123-QED}

\title{Giant exciton-phonon coupling and zero-point renormalization in hexagonal monolayer boron nitride}
\author{Himani Mishra${}^\ddagger$}
 \author{Sitangshu Bhattacharya${}^\ddagger$}
 \email{Corresponding Author's Email: sitangshu@iiita.ac.in}
\affiliation{${}^\ddagger$Nanoscale Electro-Thermal Laboratory, Department of Electronics and Communication Engineering, Indian Institute of Information Technology-Allahabad, Uttar Pradesh 211015, India}
\begin{abstract}
We report here a giant zero-point energy renormalization of 273 meV in the direct band-gap at $\textbf{K}$ in the Brillouin zone and a 571 meV of blue-shifting in the position of the doubly-degenerate brightest excitonic peak in monolayer hexagonal boron nitride. The non-radiative exciton line-width is found to be 97 meV at 0 K with a large coupling strength of 1.1 eV. This line-width is found to be mainly dominated by the scattering from the longitudinal optical phonons near the degenerate LO-TO mode, with negligible contributions from other lower branches. Additionally, the band-gap has a temperature dependent slope of -0.53 meVK$^{-1}$, which we found to be in excellent agreement with the reported experimental data on large diameter boron nitride nanotubes. We obtained our results by solving a coupled electron-hole Bethe-Salpeter equation which includes the lattice vibrational dynamics, purely using an ab-initio approach.
\end{abstract}
\maketitle

\section{\label{sec:level1}Introduction}
\noindent Bulk hexagonal boron nitride (hBN) has witnessed its controversial journey from being called a direct band-gap material \cite{Watanabe2004} to an indirect one \cite{ Arnaud2006, Cassabois2016}. The conclusion was asserted both from an ab-initio perturbative theory \cite{Arnaud2006} and two photon absorption spectroscopy \cite{Cassabois2016} that revealed a phonon assisted exciton recombination. Particularly, it was found that the photoluminescence (PL) spectrum consist of two peak regions \cite{Cassabois2016}. The first region comprised of a triplet structure between 5.75-5.80 eV, that demonstrated an emission process assisted by the LO-TO (longitudinal and transverse respectively) optical phonon branch along the $\Gamma$-$\textbf{K}$ direction in the Brillouin zone (BZ) \cite{Cassabois2016}. The second region contains a doublet structure between 5.85-5.90 eV that displayed an absorption process communicated by the LA-TA acoustic branch along the same direction. Theoretical support to confirm this phonon assisted luminescence evolved recently by calibrating a non-equilibrium Green's function to solve the exciton dynamics taking the first order phonon perturbative theory and real-time Bethe-Salpeter equation (BSE) \cite{Cannuccia2018}. Infact, a static BSE approach was also solved earlier \cite{Marini2008} to get the non-radiative line-width of the excitons in the presence of lattice vibrations.\\
With an indirect band-gap of 5.96 eV and a single particle gap of 6.08 eV \cite{Cassabois2016}, high quality crystalline bulk hBN gathered incredible popularity owing to its extraordinary luminescence in the deep ultraviolet region. It has been recently found that due to very high oscillator strength, this luminescence efficiency can be as large as 15$\%$ compared to only 0.1$\%$ offered by diamond \cite{Schue2018}. Furthermore, it was also found that due to the interference of two groups of transition with opposite signs at the two-fold valley degrees of freedom \textbf{K} and $\textbf{K}^{\prime}$ in the BZ, the loss function at finite exchange momentum \textbf{q} along $\Gamma$-$\textbf{K}$ also exhibits strong intensity \cite{Sponza2018a}. Infact, there also has been many speculations about the nature of excitons in this material. It is much under debate about whether one should call this lowest-lying bright bound exciton as a Wannier type \cite{Watanabe2004, Sponza2018b}, Frenkel type \cite{Arnaud2006}, quasi-Frenkel type or simply tightly bounded \cite{Cannuccia2018, Schue2018}. These claims have already stirred up the curiosity to unveil exciton dynamics in monolayer (ML) hBN.\\
Strategic understanding of excitonic behaviour in ML hBN has recently surfaced by (a) altering the bulk stacking and weak van der Waals inter-layer distances \cite{Aggoune2018, Sponza2018b, Koskelo2017} unfolding the spatial distribution of the two-dimensional (2D) exciton, (b) revealing of exciton band structure at finite \textbf{q} to characterize exciton wave-function \cite{Cudazzo2016}, (c) Davydov splittings \cite{Paleari2018}, (d) dimensional dependencies \cite{Wirtz2006, Golberg2010}, etc. In addition, equivalent tight binding formalism \cite{Galvani2016} is also in the assembly line to simplify the complicated and computationally demanding perturbative absorption theory.\\
As experiments are performed at finite temperature T, it stands therefore a bottle-neck problem to realize the temperature dependent absorption spectra. It appears that all the above perturbative work done on ML hBN assumed a frozen-atom condition (i.e., atoms are assumed to be located fixed in their equilibrium lattice points), neglecting electron-phonon interactions which causes the spectra to be temperature dependent and consequently broadens the exciton lifetime. Therefore, in this work we try to respond to the following two crucial questions: (a) how does the electron-phonon interactions renormalizes the ML hBN band-gap? and hence, (b) how does the temperature controls the corresponding absorption spectra, excitonic energies and the non-radiative line-widths? Indeed, performing an exhaustive calculation of the phonon perturbative self-energies- the Fan and Debye-Waller \cite{Fan1950, Elena2011, Marini2008} respectively, we found two appealing situations at 0 K:
\setlist{nolistsep}
\begin{itemize}[leftmargin=*]
\item the band-gap shrinks down by a giant 273 meV from its corresponding value at the frozen atom condition and shrink further as T rises, and
\item the absorption peak of the lowest bright exciton is blue-shifted by about 571 meV from its corresponding value at the frozen atom condition. The peak continuously red-shifts as T increase.
\end{itemize}
\noindent
We unfold that the root cause of the brightest excitonic non-radiative line-width is the scattering from the LO phonons. We confirm this using two methods, namely by exploiting the Eliashberg function \cite{Mahan2014} from the aforementioned self-energies as well as by showing an exponential dependency of the excitonic line-width on temperature. We also demonstrate a temperature dependent band-gap with a slope of -0.53 meVK$^{-1}$. We found this to be in excellent agreement with the reported experimental data on large diameter boron nitride nanotubes \cite{Du2014}.\\ 
Our methodology is based on fully state-of-the-art ab-initio approach involving a converged density functional theory (DFT), density functional perturbation theory (DFPT) and many-body perturbation theory (MBPT) calculations respectively and is free from any fitting parameter. What follows, in Section II we outline our computational methodology to achieve the results. This is followed by the results and discussions section III which outlines the main outcomes like the ground state and excited state energies, the electron-phonon interaction strengths and band-gap modification along with extraction of finite T-absorption spectra, line-widths and exciton energies. Section IV summarizes our results. All the mathematical formulations are categorized in the Appendix section. Additionally, the supplementary information \cite{Supplemental} contains supportive figures and convergence criteria.
\begin{figure*}[!ht]
\includegraphics[width=1.8\columnwidth]{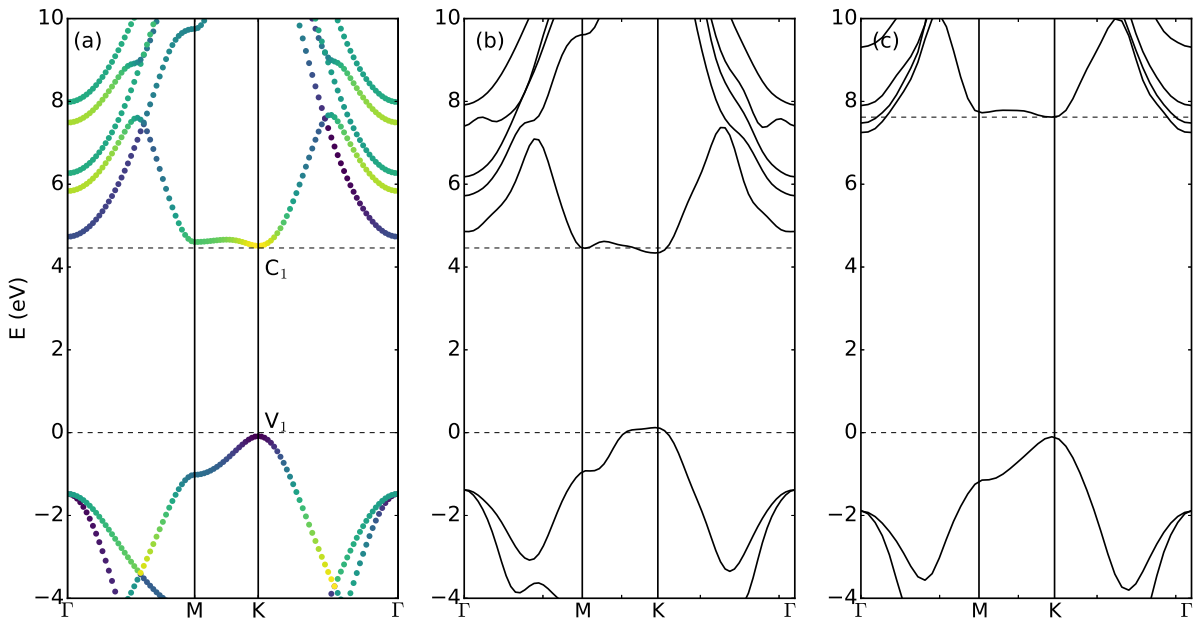}
\caption{(a) Frozen-atom ground state band-structure of ML hBN along the high symmetry BZ. The band-gap is direct at \textbf{K}. $\mathrm{C}_{1}$ and $\mathrm{V}_{1}$ represent the bottom of conduction and top of valence band at \textbf{K} respectively. The color variation exhibits orbital density-of-states contribution of boron and nitrogen atoms projected on the bands. A strong violet to yellow color change shows nitrogen to boron character. (b) Band-structure at 300 K. The band-gap is still direct at \textbf{K} but with a reduction of 370 meV due to electron-LO phonon energy renormalization. (c) Single shot GW (G$_{0}$W$_{0}$) band-structure under frozen atom condition exhibiting a direct band-gap of 7.73 eV at \textbf{K}. The indirect gap between the bottom of the conduction band at $\Gamma$ and top of the valence band at \textbf{K} ($\Gamma$-\textbf{K}) is 7.32 eV.}
\end{figure*}
\section{\label{sec:level1}Computational Details}
\paragraph{Ground state calculations:} Bulk hBN possess D$_{4h}$ ($\overline{6}$2m) symmorphic group symmetry which reduces to D$_{3h}$ ($\overline{6}$2m) in case of ML sheet. A unit cell consisting of 2 atoms was created with a vacuum-slab-vacuum profile of 16$\mathring{\mathrm{A}}$ to minimize the Coulombic interactions between the repeated images. A kinetic cut-off energy of 120 Ry (See Fig. S1 \cite{Supplemental} for convergence with respect to kinetic cut-off energy) and a norm-conserving pseudo-potential with an exchange-correlation functional at the level of local density approximation (LDA) were chosen. As boron and nitrogen are both light elements, spin-orbit interactions were found negligible and therefore neglected \footnote{We did not find any spin-orbit splitting even using a fully relativistic and norm-conserving pseudo-potential at the level of LDA (see Fig. S8 in \cite{Supplemental}). To the best of our knowledge we yet found no experimental evidence, neither in the bulk nor in monolayers of hBN, which signature spin-splitting in the peaks of the absorption spectrum. Instead, the effect of spin-orbit interactions on exciton-phonon coupling and zero-point energy are much pronounced in similar structures of TMDC families like WSe$_{2}$, MoS$_{2}$, etc. where the conduction and valence band splitting are found to be relatively quite large \cite{He2014, Molina2016}. In such cases the peaks in the absorption spectra are also spin-splitted.}. A $\Gamma$ centred Monkhorst-Pack scheme was used to sample the BZ on a 12$\times$12$\times$1 grid. The DFT package $\mathrm{\mathsf{\mathtt{Quantum}}}$ $\mathrm{\mathsf{\mathtt{Espresso}}}$ \cite{Giannozzi2017} code was then used to solve the ground state Kohn-Sham energies. This resulted in a lattice constant of 2.48 $\mathring{\mathrm{A}}$ after reducing the force and energy thresholds down to $10^{-5}$ Ry/Bohr and $10^{-5}$ Ry respectively. A convergence test on BZ sampling upto 36$\times$36$\times$1 grid shows insignificant gap improvement (See Fig. S2 \cite{Supplemental} for convergence on k-points).
\paragraph{Electron-phonon self-energies:} A uniform 12 $\times$ 12 $\times$ 1 dense phonon grid was found sufficient to calculate the phonon dispersion along the BZ. DFPT computations were now used to calculate the first and second order electron-phonon matrix elements. These corrections were done for all the electronic valence bands and 11 lowest conduction bands. To achieve this, the irreducible BZ was sampled with a dense 200-random irreducible BZ \textbf{q}-points. It should be noted that a rather uniform phonon grid can also be used to extract these self-energies, however the use of random grid provides an additional advantage of assigning an implicit weight to electron-phonon self-energy integral. We find that in literature, the general consensus is to use random \textbf{q}-grids for the evaluation of the electron-phonon self-energies \cite{Elena2011, Molina2016, Felician2010}.
\paragraph{Excited-state G$_{0}$W$_{0}$ and BSE calculations:} The excited state energies and spectra, i.e., the Hedin's GW \cite{Hedin1965} and BSE computations were performed with the MBPT package YAMBO code \cite{Andrea2009}. The dynamic dielectric screening function in the G$_{0}$W$_{0}$ calculations was evaluated with a generalized Godby-Needs plasmon-pole approximation model \cite{Godby1989}. 100 bands (4 occupied and 96 unoccupied) were considered in the local field effects calculation for the determination of the polarization function within the random phase approximation (RPA) screening level with a response block size of 500 reciprocal-lattice vectors. This is an energy cut-off of about 13 Ry. Additionally, 20065 \textbf{G}-vectors (equivalent to almost 150 Ry) were used to expand the wave-function in the plane wave basis set for the calculation of Hartree-Fock exchange self-energy considering the same 100 electronic bands and summed-up to get the convergence. One of the major challenges when dealing with two-dimensional (2D) systems is the finite length in one of the spatial direction. This introduces rapid variations in screening and as a result the integral quantities like exchange self-energies, BS kernel, total energy expression, etc. suffers \textbf{q}$\rightarrow$0 divergence problem due to the quasi-2D nature of Coulomb interaction. In order to compute those quantities properly, random integration method provides a robust methodology \cite{Olivia1998, Andrea2009, Rozzi2006}. This method assumes a smooth momenta integrand function centred on \textbf{q} in each small volumetric region of the BZ without disturbing the potential itself. A Montecarlo method was then used to evaluate this BZ volume integral which forbids the divergence to happen. The same method was used to fix the divergences in the electron-phonon self-energy calculations as well as in the BSE. Such methods are extremely robust and used more often whenever integral divergence appears. We used 10$^{6}$ random \textbf{q}-points which we found to cover the BZ fully and is employed for the first 111 \textbf{G}-vectors (i.e., a cut-off of 4 Ry) of the Coulomb potential to extract the self-energies associated in the G$_{0}$W$_{0}$ calculations. We find that this much number of random points and \textbf{G}-components are sufficient for a converged solution using the random integration method. Additionally, 32 $\mathring{\mathrm{A}}$ on either side of the ML was used in a form of box-structure to truncate the Coulomb potential between repeated images \cite{Rozzi2006}. \\
Optical spectra calculations are extremely delicate and depend on the way the entire BZ is sampled. To speed up the numerical computations at very dense grid without losing the numerical accuracies, we use the methodologies presented by Kammerlander, et. al. \cite{Kammerlander2012}.  We first dense sample the BZ to a shifted 24$\times$24$\times$1 grid and then map it to the corresponding 12$\times$12$\times$1 grid states calculations. First the BSE is solved taking the independent particle approximation in a shifted dense grid using a Wannier interpolator. This is followed by again solving the interacting BS kernel in the 12$\times$12$\times$1 grid keeping the contributions from the fast changing independent-particle results. This is a very useful technique that minimizes the computational burden to a large extent keeping the same level of numerical accuracies \cite{Alejandro2013, Molina2016, Kammerlander2012}. To build the interacting BS kernel, the static dielectric screening was first computed within the RPA (i.e., Hartree potential only) using the same 100 electronic bands. In order to see the effect of electron-phonon interactions on the energy bands, a scissor operator of 3.129 eV was applied to mimic the G$_{0}$W$_{0}$ gap. Along with this, a linearly fitted stretching factor of 1.07 eV and 1.26 eV was imposed to the conduction and valence bands respectively (See Fig. S3 \cite{Supplemental} for QP corrections over the corresponding LDA energies). In the frozen atom approximation, we include only the QP energies directly from the G$_{0}$W$_{0}$ calculations in the BS kernel (G$_{0}$W$_{0}$+BSE). The light-polarization vector direction was included in the local field effects and kept parallel and perpendicular to the hBN monolayer plane to see its effect on absorption spectra. Additionally, we go beyond the Tamm-Dancoff approximation \cite{Fetter2013, Rohlfing2000} to include both the resonant and anti-resonant electron-hole pairs in the BS kernel. This captures properly the collective density oscillations from electron-hole pairs. Along with this, all the valence bands and 4 lowest conduction bands were used to get the converged absorption spectra (See Fig. S4(a-b) \cite{Supplemental} for the convergences with respect to sampling and bands respectively). The BSE hamiltoninan was then finally diagonalized to get the excitonic states and eigenvalues.
\section{Results and Discussions}
\begin{figure*}[!ht]
\includegraphics[width=1.9\columnwidth]{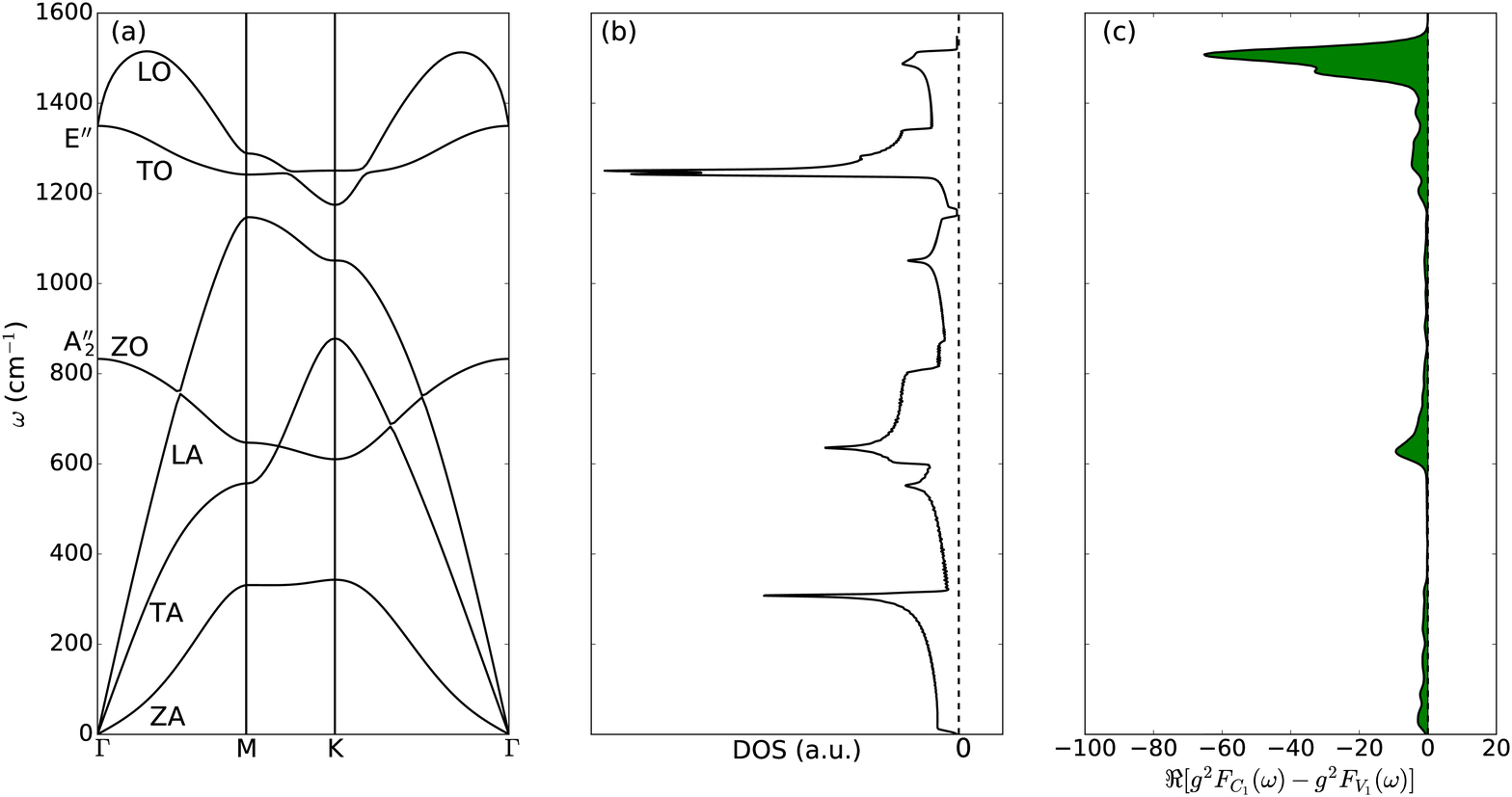}
\caption{(a) Phonon dispersion of ML hBN along high symmetry BZ. The LO-TO branch at $\Gamma$ is degenerate at 1380 cm$^{-1}$. The out-of-plane acoustic ZA mode is parabolic in phonon wave-vector. (b) Phonon density of states as function of lattice vibrational frequencies. (c) The real part of the Eliashberg function difference ($\Re\left[g^{2}F_{\mathrm{C_{1}}}\left(\omega\right)-g^{2}F_{\mathrm{V_{1}}}\left(\omega\right)\right]$) as function of lattice vibrational frequencies between conduction and valence bands at \textbf{K}. The most prominent peak at 1529 cm$^{-1}$ signals the LO branch involvement in processes like an exponential increment in exciton line-widths.}
\end{figure*}
\noindent Bulk hBN is a wide band-gap material and therefore, its ML sheet is expected to have even larger gap. Fig. 1(a) exhibits the frozen-atom ground state electronic energies along the high symmetry BZ with a bare direct-gap of 4.596 eV at \textbf{K}, consistent with others \cite{Kang2016, Rasmussen2016, Berseneva2013, Galvani2016}. The direct gap is mainly formed by the boron ($\pi^{*}$-band, hole distribution) and nitrogen ($\pi$-band, electron distribution) $p_{z}$-orbitals. This is demonstrated by projecting the orbital contributions on the band energies in which a strong violet (yellowish) color exhibits a nitrogen (boron) character respectively. The orbital weightage projected on each atom locally is exhibited in the supplementary Fig. (S5) \cite{Supplemental}.
\begin{figure*}[!ht]
\includegraphics[width=1.8\columnwidth]{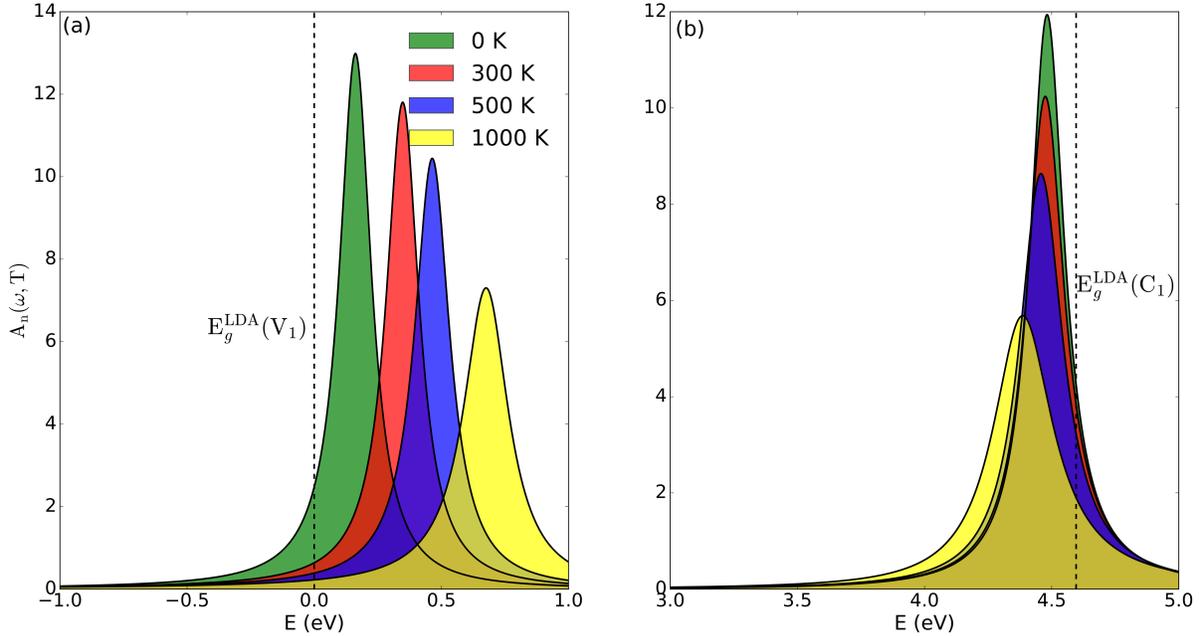}
\caption{Spectral function at different temperatures for (a) top valence band and (b) bottom conduction band at \textbf{K} in the BZ (See Fig. 1(a)). As the temperature increases, the valence peak blue-shifts while the conduction peak red-shifts exhibiting a reduction in the gap. The 0 K gap is 273 meV less than that offered by the frozen atom condition.}
\end{figure*}
Single-shot GW corrections on these bare energies were found sufficient to open the gap upto 7.73 eV at \textbf{K} and is found to be in agreement with Rasmussen et. $al$. \cite{Rasmussen2016}. The band-gap now became an indirect between the bottom of the conduction band at $\Gamma$ and top of the valence band at \textbf{K} ($\Gamma$-\textbf{K}) with a value of 7.32 eV. This is demonstrated in Fig. 1(c). This change of direct gap (as a consequence of frozen atom DFT calculation) to an indirect gap (G$_{0}$W$_{0}$ computation) has also been previously reported \cite{Blase1995, Galvani2016}. In fact, Blase et. $al.$ \cite{Blase1995} demonstrated that within the LDA level, the maximum charge density associated with the lowest unoccupied state at ${\Gamma}$ is located at 3.3 a.u. away from the in-plane atoms. This location is far compared to the maximum charge density associated with the $p_{z}$-orbitals which is located only at 0.75 a.u. away from the in-plane atoms. Thus the lowest unoccupied state at ${\Gamma}$ behaves like a nearly free electron character which results in a lowering of the electronic energy at $\Gamma$ point. Additionally, the real part of the QP renormalized weight factor ($Z$) of conduction and valence bands at \textbf{K} was found to be 82$\%$ and 85$\%$ respectively. These values are comparable to the sharp QP states shown for similar hexagonal ML families and others \cite{Elena2011, Cannuccia2012, Mishra2018}.\\
\begin{figure*}[!ht]
\includegraphics[width=1.8\columnwidth]{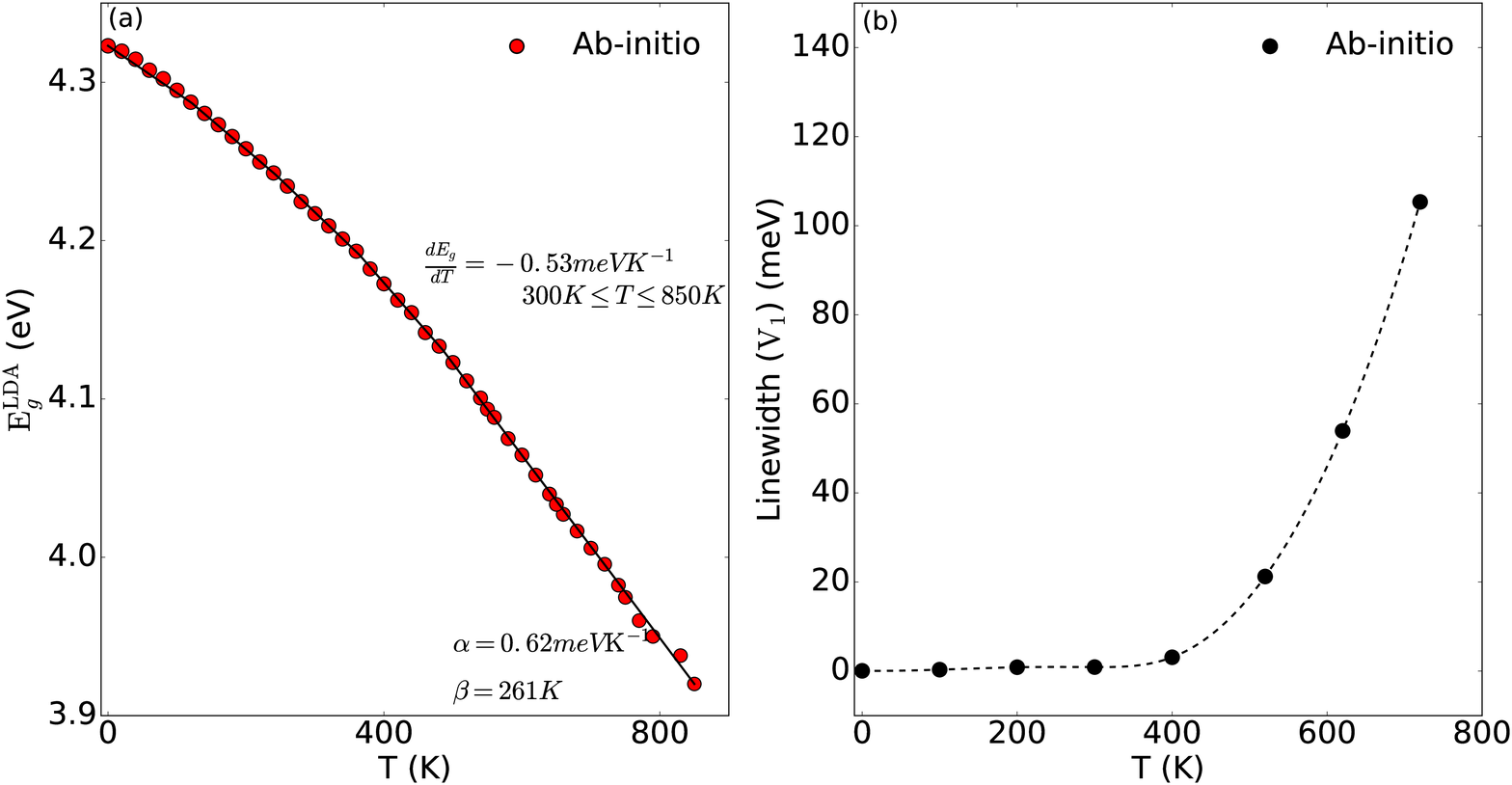}
\caption{(a) Band-gap as a function of temperature with a slope of -0.53 meVK$^{-1}$. (b) Valence band line-width as function of temperature demonstrating an exponential dependency due to electron-LO phonon interaction. Symbols are the ab-initio data while the lines are the guide to an eye.}
\end{figure*}
\noindent The lattice vibrational results in terms of phonon dispersions and density of states has been shown in Fig. 2(a-b) respectively. Since hBN is a polar material, the long-range Coulombic interaction and electronic screening were expected to modify the optical phonon dispersion thereby splitting the LO-TO branch at $\Gamma$. ML hBN seems to breakdown this convention \cite{Sohier2017}. Keeping this in view, the interatomic force constants were carefully calculated following \cite{Sohier2017} in real space from the dynamical matrices such that the norm of the phonon momentum vanishes along with the Born-effective charges determining the slopes. The zone centre frequencies in Fig. 2(a) were thus found to consist an infra-red active optical out-of-plane ($A_{2}^{\prime\prime}$) (ZO) mode almost at 848 cm$^{-1}$ and two degenerate LO-TO infra-red and Raman active ($E^{\prime}$) mode at 1380 cm$^{-1}$. The out-of-plane vibrational acoustic mode (ZA) exhibits a quadratic behaviour over the BZ path and provides a clear signature of a non-zero ZA phonon density-of-states when $\omega \rightarrow$0. Indirectly this also confirms that in our present work, the hBN interlayer atomic plane separations are quite large. Using the 200 irreducible random \textbf{q}-points, the dynamical matrices were constructed to calculate the frequency-dependent Fan and the frequency-independent Debye-Waller self-energies (See Appendix for mathematical description). It should be noted that the former stems out from the spatial charge density variations, while the latter is due to the translational invariance of the crystal. From these self-energies, the pole of the single particle Green's propagator $G_{n,\textbf{k}}\left(\omega\right)$ at each state $\left|n,\textbf{k}\right\rangle$ was constructed whose real part is the renormalized QP energy while the imaginary part is the lifetime. This is known as the dynamical Heine, Allen and Cardona or HAC theory \cite{Fan1950, Elena2011, Marini2008, Allen1983, Allen1976} which leads to a non-zero electronic energy even when T$\rightarrow$0. This residual energy is also known as the zero-point energy and is in accordance with the Heisenberg's uncertainty principle. The static condition (the so-called on-the-mass-shell approach \cite{Marini2008}) of this theory assumes both the real and imaginary Fan self-energies to be independent of frequency when the bare energies lie far from the poles \cite{Cannuccia2011, Antonius2015}. We rather use a more general dynamical theory in our calculations which neglects the static condition (See the Appendix for a mathematical approach). The imaginary part of the Green's function can also be represented as the spectral function $\mathrm{A}_{n\textbf{k}}\left(\omega\right)=\frac{1}{\pi}\left|\Im G_{n\textbf{k}}\left(\omega\right)\right|$ and is shown in the Fig. 3(a-b) for both top of the valence band ($\mathrm{V}_{1}$) and bottom of the conduction band ($\mathrm{C}_{1}$) at $\textbf{K}$ (See Fig. 1(a)). The inclusion of lattice vibrations is then immediately clear which demonstrates a zero-point energy of 161 meV for top of the occupied band (i.e., $\mathrm{V}_{1}$ crosses 0 eV level and moves up by 161 meV) and 112 meV for bottom of the unoccupied band (i.e., $\mathrm{C}_{1}$ crosses 4.596 eV level and moves down by 161 meV) from their respective bare energies, resulting in a giant 273 meV shrinking of the band-gap. Fig. 1(b) shows this effect at 300 K, where the band-gap reduction becomes as high as 370 meV. These remarkable energies cannot be obtained from the standard ground state DFT calculations. As the temperature is increased, the peaks of the spectral function become dwarf and acquire asymmetric Lorentzian shapes with reduced QP renormalized weight factors signifying intense electron-phonon interaction strength. The QP energy difference between the two peaks (for example, the top valence and bottom conduction at \textbf{K}) also reduces, confirming a band-gap reduction with temperature. It is then interesting to calculate the slope $\frac{dE_{g}}{dT}$ from the band-gap $E_{g}$. We found this to be -0.53 meVK$^{-1}$ as exhibited in Fig. 4(a). This is in excellent agreement with the experimental slope of -0.43 meVK$^{-1}$ measured for BN nanotubes of diameter 60 nm \cite{Du2014}. We used the Varshni's equation \cite{Varshni1967} to fit the temperature dependent band-gap from $E_{g}\left(T\right)=E_{g}\left(0K\right)-\frac{\alpha T^{2}}{\left(T+\beta\right)}$ in which $\alpha$ and $\beta$ are the Varshni coefficients. We found that $\alpha$=0.62 meVK$^{-1}$ and $\beta$=261 K in our present case. It should be noted that we have chosen the temperature not to go beyond 850 K. This is because free-standing hBN monolayer can withstand temperature up to 850 \textdegree{C} (1123.15 K) \cite{Hua2014}. Near this temperature regime our methodology may not properly capture the associated phase-transition \cite{ Cannuccia2012}. Rather, we found that beyond 850 K, the band-gap exhibit anomalous behaviour. A careful analysis exhibit that this could be due to a large reduction in the real part of the QP renormalization factor (Z$\sim$33$\%$). Such low Z value is questionable for a good QP state \cite{ Cannuccia2012} and therefore our calculated slope is valid between 300 K to 850 K.\\
\begin{figure*}[!ht]
\includegraphics[width=1.8\columnwidth]{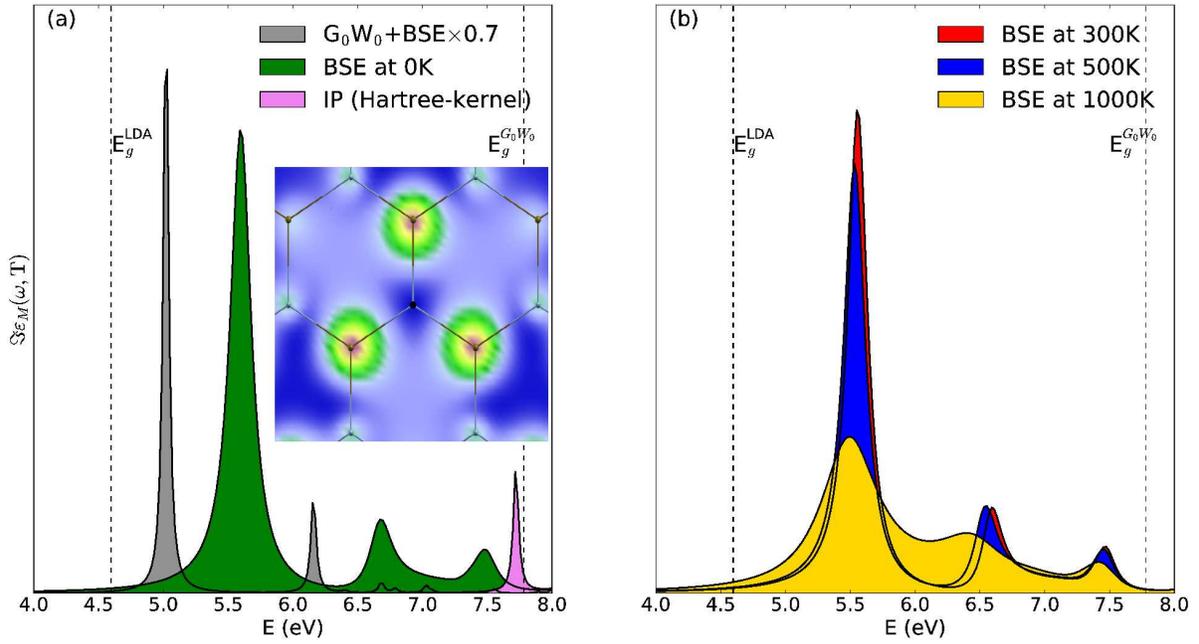}
\caption{(a) Absorption spectra under frozen atom G$_{0}$W$_{0}$+BSE (grey),  under frozen atom independent-particle approximation (magenta) and at T = 0 K (green). The kernel in the G$_{0}$W$_{0}$+BSE spectra (scaled down by 30 $\%$) is the sum of a bare positive Coulomb exchange term (repulsive) and a direct negative screened electron-hole interaction (attractive). The kernel in the independent particle approximation is allowed to contain upto the Hartree potential. The inset shows the excitonic wave-function unfolded over the ML hBN lattice at 300 K. The dark spot represents the hole over nitrogen atom at a distance of 1 $\mathring{\mathrm{A}}$. (b) Absorption spectra at various other temperatures. All the peaks red-shift as the temperature rises. This is further shown in the inset where the lowest bright exciton peak position decreases with temperature. Symbols are the ab-initio data while lines are the guide to an eye.}
\end{figure*}
\begin{figure*}[!ht]
\includegraphics[width=0.9\columnwidth]{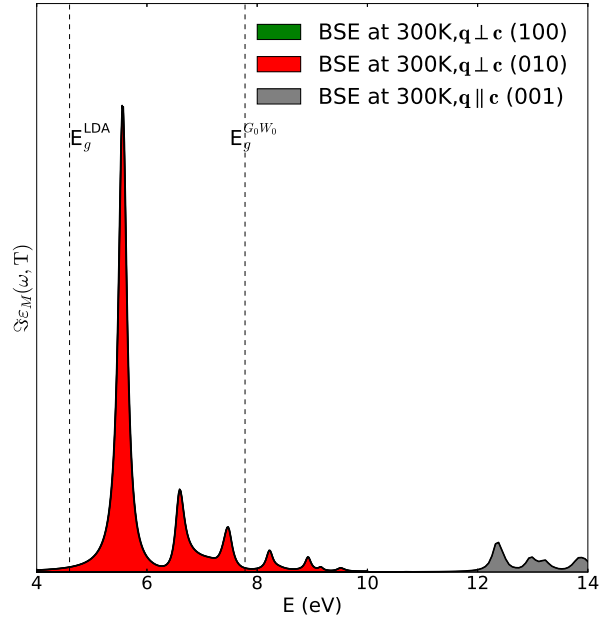}
\caption{Anisotropic absorption spectra with the light polarization direction along the \textbf{c}-direction at 300 K. For comparison, we put the corresponding in-plane spectra also. A strongly quenched and blue-shifted spectra is found which is due to the charge inhomogeneity along the \textbf{c}-direction with contribution from $p_{z}$ orbital.}
\end{figure*} 
\begin{figure*}[!ht]
\includegraphics[width=0.85\columnwidth]{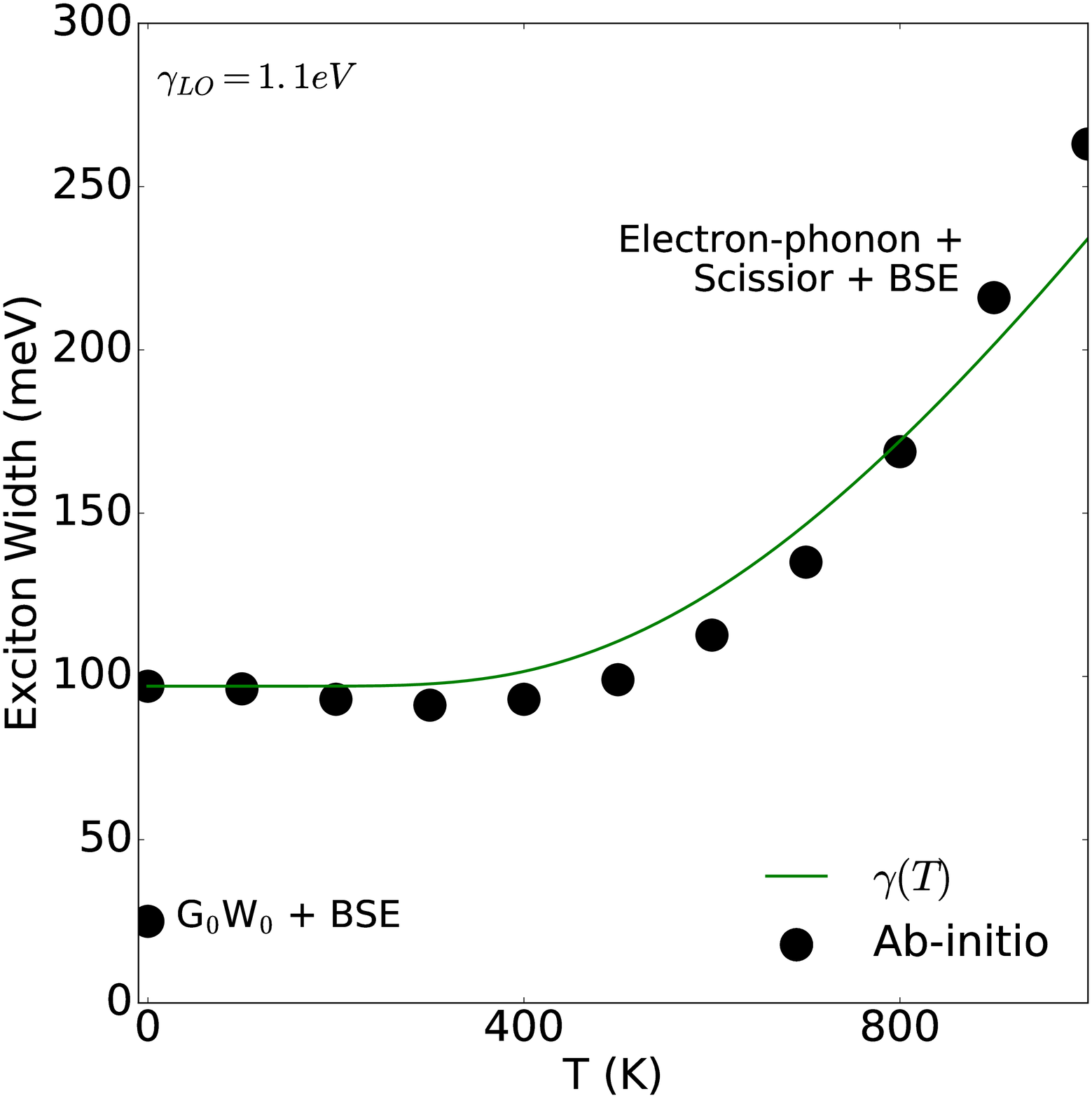}
\caption{Lowest bright exciton line-width as function of temperature exhibiting an exponential dependency as temperature rises. The 0 K width is 97 meV, almost thrice compared to the corresponding frozen atom G$_{0}$W$_{0}$+BSE case. A giant exciton-LO-phonon coupling constant of 1.1 eV was found. Symbols are the ab-initio data while line corresponds to $\gamma(T)$.}
\end{figure*}
\noindent It is worthy to note now that the band-gap also reduces due to the presence of lattice anharmonicity. The lattice anharmonic contribution causes the lattice thermal expansion, constraining the band-gap to reduce further. However, we do not intend to compute this using a standard quasi-harmonic approximation approach \cite{Mounet2005}, as it is well-known that unless there is no portion of the real part of the Eliashberg function difference taken between the lowest conduction and the top valence band (i.e, $\Re\left[g^{2}F_{\mathrm{C_{1}}}\left(\omega\right)-g^{2}F_{\mathrm{V_{1}}}\left(\omega\right)\right]$) that acquires a positive value, $E_{g}\left(T\right)$ will always be much larger than $E_{g}^{LTE}$ \cite{Mahan2014, Villegas2016, Tsang1971}. This is also an indirect proof to predict whether the gap decreases with T or not. Such anomalous behaviour (i.e., gap increases with T) has been observed in some special cases like black phosphorous \cite{Villegas2016}, but in case of ML hBN we found this difference $\Re\left[g^{2}F_{\mathrm{C_{1}}}\left(\omega\right)-g^{2}F_{\mathrm{V_{1}}}\left(\omega\right)\right]$ to remain always negative till the Debye frequency is reached. This is particularly displayed in Fig. 2(c). Also the difference $\Re\left[g^{2}F_{\mathrm{C_{1}}}\left(\omega\right)-g^{2}F_{\mathrm{V_{1}}}\left(\omega\right)\right]$ $\rightarrow$0 as $\omega\rightarrow0$, implies the conservation of the crystal translational invariance property. The renormalized line-widths at \textbf{K} for $\mathrm{V}_{1}$ are shown in Fig. 4(b). A careful observation exhibits that the variation is exponential at high temperatures indicating the involvement of LO phonons at 1529 cm$^{-1}$. We now discuss the implication of this on the broadening of the electronic band-structure. The uncertainty principle dictates that lifetimes are inversely proportional to the line-widths. Thus a smaller line-width would mean that the particle spends more time in that state. However as the temperature rises, the electron-phonon (in this case electron-LO phonon) interaction become more intense (i.e., the line-width increases) and as a result the electron very quickly gets scattered to a new state. This is the reason why the top of the valence band at \textbf{K} should possess a small but finite line-width at 0 K .\\
A coupled electron-hole BSE calculation in the optical limit (\textbf{q}$\rightarrow$0) was carried out at different temperatures to obtain the absorption spectra. The electron-phonon line-widths from the aforementioned results were included in the BSE Hamiltonian. This however makes the Hamiltonian to be a non-hermitian. Thus a full diagonalization technique is necessary to get the exciton line-widths, weights, amplitudes and wave-functions. Figure 5(a-b) exhibits the spectra for three cases namely, the single shot GW+BSE, scissor correction+BSE+polaronic widths and the independent particle absorption (containing upto the RPA correction in the BS kernel). In all these cases, the light polarization direction was kept $\perp$ to the \textbf{c}-direction. The single shot GW+BSE (in absence of lattice vibrations) gives the first excitonic peak position at 5.02 eV leading to a giant exciton binding energy (BE) of 2.71 eV. This peak is due to the transition of electrons from $\mathrm{V}_{1}$ to $\mathrm{C}_{1}$ at \textbf{K}. In order to get the usual Lorentzian spectra shape, an additional broadening of 0.1 meV is used in this case. This makes the peak narrow and sharp, making the lifetime very large (infinite, if additional broadening would be 0).\\ 
Immediately a finite excitonic lifetime appears when the electron-phonon renormalized energies and corresponding line-widths are incorporated in the BSE hamiltonian. The corresponding spectra is shown for T = 0 K, where the manual broadening of 0.1 eV is now no longer required. The effects of polaronic widths are indeed large on the first bright exciton peak. This can be seen through a 571 meV of blue-shifting from the case of G$_{0}$W$_{0}$+BSE. The peak is now at 5.59 eV resulting in the giant BE of 2.14 eV. Strictly speaking, the IP approximation is the special case when the BSE kernel is zero and thus, the dielectric response function does not contain any electron-hole interaction. However in the present case, we limit the BSE kernel to only upto the Hartree potential (i.e., the RPA only). This is the reason why the peak nearly at 7.5 eV is just below the G$_{0}$W$_{0}$ gap. This situation thus grossly overestimates the absorption peak position. Figure 5(b) exhibits the effect of three additional temperatures on the spectra. We see that the spectra slowly diminish with temperature, while the line-width significantly broadens by the time T reaches 1000 K. The peaks also red shift, resulting in the exciton energy reduction governed by the Varshni equation or the phenomenological relation \cite{Donnell1991} when excitons are present. The 300 K brightest peak position is found to be at 5.55 eV which red shifts to 5.49 eV at 1000 K. To understand which phonon mode has caused the exciton line-width to broaden can be unveiled by the Eliashberg function evaluated at each $\left|n,\textbf{k}\right\rangle$. Since the main absorption peak is due to the electronic transition which occurs from $\mathrm{V}_{1}$ to $\mathrm{C}_{1}$ at \textbf{K}, we calculate the quantity $\Re\left[g^{2}F_{\mathrm{C_{1}}}\left(\omega\right)-g^{2}F_{\mathrm{V_{1}}}\left(\omega\right)\right]$ at these two states. A strong peak at 1529 cm$^{-1}$ (See Fig. 2(c)), which is at the LO phonon branch, is thus found to be responsible for the exciton scattering. Although a similar and opposite case was found with MLs like MoS$_{2}$ \cite{Molina2016} and WSe$_{2}$ \cite{Mishra2018} respectively, however the underlying reasons were entirely different due to the involvement of strong spin-orbit couplings. We found here that at all temperatures the lowest exciton remains bright and degenerate with a dark exciton. This is in stark difference to the case with excitons at \textbf{q}$\rightarrow$0 in bulk-hBN. At such condition, the doubly degenerate dark exciton house themselves in lower energy compared to the doubly degenerate bright excitons \cite{Koskelo2017, Paleari2018}. This degeneracy between bright and dark is removed at finite \textbf{q} where the later possess a very small oscillator strength \cite{Cudazzo2016}. The inset in Fig. 5(a) shows the excitonic wave-function of the lowest lying doubly-degenerate exciton at 300 K plotted over the in-plane coordinates. In order to find out the excitonic probability density, we have fixed the hole above the nitrogen atom in the atomic plane at a distance of 1 $\mathring{\mathrm{A}}$. This fixation has been done in order to account for the fact that the valence band is mainly formed by the $p_{z}$ orbital of nitrogen atom at \textbf{K} in the BZ as shown in Fig. 1 (a). We see that the exciton wave-function is strongly localized around the hole over the boron atoms and exhibits a triangular shape preserving the triangular symmetry of the lattice \cite{Wirtz2008}. Further, there is a slight contribution to the wave-function also from the nearest neighbours comprising boron atoms. This implies that both boron and nitrogen orbitals have contributions in the excitonic wave-function build-up. The corresponding wave-function in the bulk case is found to be almost same \cite{Cudazzo2016} as in this present ML hBN. This is due to the reason that the excitonic wave-function in the former is also confined in the single layer. The inset in Fig. 5 (b) demonstrates a decreasing trend of exciton peak energy position with temperature which is evident from an increasing line-width as well as decreasing optical band-gap. The trend follows the well-known relation \cite{Donnell1991} $E_{p}\left(T\right)=E_{p}\left(0K\right)-S\left\langle \hbar\omega\right\rangle \left[\mathrm{coth}\frac{\left\langle \hbar\omega\right\rangle }{2k_{B}T}-1\right]$, in which $S\left\langle \hbar\omega\right\rangle $ is the dimensionless coupling constant dependent on the average phonon energy $\left\langle \hbar\omega\right\rangle $. We find that $S\left\langle \hbar\omega\right\rangle $=27.67. This is one order larger than the ML transitional metal-dichalcogenides (TMDC) \cite{Cadiz2017} exhibiting a strong exciton binding energy.\\
In Fig. (6), we have demonstrated the absorption of in-plane as well as out-of plane linearly polarized light at 300 K. The out-of-plane absorption is strongly blue shifted with respect to the in-plane absorption. This is because the light emission in ML hBN arises mainly due to the transitions from valence band maxima to conduction band minima with a $p_{z}$ orbital character. An intense charge anisotropy present along the hBN-vacuum interface further results in a depolarization effect, i.e., a strong quenching in absorption. The excitonic line-widths as function of temperature have been shown in Fig. 7. These line-widths are the non-radiative ones whose rate of variation signals the exciton-phonon scattering strength. We thus use the phenomenological relation between the phonon induced exciton line-width and temperature \cite{Selig2016} $\gamma\left(T\right)=\gamma\left(0K\right)+\gamma_{ac}T+\gamma_{op}\left[\textrm{exp}\left(\frac{\hbar\omega_{op}}{k_{B}T}\right)-1\right]^{-1}$ in order to characterize such strengths.  We find a negligible contribution from the small peak near 600 cm$^{-1}$ (See Fig. 2(c)) which gives an almost zero contribution from the acoustic phonon dominated linearly dependent term. We obtained a giant $\gamma_{op}$=1.1 eV as the exciton-LO phonon strength signifying an exponential variation. The line-width for G$_{0}$W$_{0}$+BSE (frozen atom) and 0 K calculations are about 23 meV and 97 meV respectively, quite larger than few tens of meVs for ML TMDC \cite{Cadiz2017}.\\
Finally, our results are yet to be confirmed by the experiments done on free-standing 2D hBN. As a bench-mark, we demonstrate that the band-gap dependency on temperature is in excellent agreement with that reported for large diameter BN nanotubes \cite{Du2014}. In all our methodologies, we did not include any defects, structural variations, exciton-exciton and other similar interactions which may impose immense computational complexities at the present stage and are beyond the scope of this work.
\section{Conclusions}              
\noindent Summarizing, we have computed the temperature dependent absorption spectra in ML hBN. We found that the lowest bright exciton is scattered by the LO phonon branch with a giant strength of 1.1 eV. The band-gap shrinks to a giant 273 meV at 0 K from its corresponding value at frozen atom condition and continues to shrink as T rises with a rate of -0.53 eVK$^{-1}$. Our computational methodology is based on both density functional theory and density functional perturbation theory. The excited state single-shot GW and coupled electron-hole Bethe-Salpeter equation are computed from the many body perturbation theory. The results presented here are purely ab-initio without any fitting parameter and are in excellent agreement with reported experimental data on large diameter BN nanotubes \cite{Du2014}.
\begin{acknowledgements}
\noindent SB acknowledges the financial support from DST, India with the grant number YSS/2015/000985 under Fast-Track Young Scientist Programme. HM acknowledges MHRD and DST, Govt. of India, for providing fellowship. Both the authors acknowledge the extra computational support from Institute's Central Computational Facility (CCF).
\end{acknowledgements}
\appendix*
\section{}    
\noindent In the  presence of lattice vibrations, the first order electron-phonon matrix elements can be written as \cite{Fan1950}
\begin{eqnarray}
g_{n^{\prime}n\textbf{k}}^{\textbf{q}\lambda}=\sum_{\alpha s}\left\langle n,\textbf{k}\left|\nabla_{\alpha s}\phi_{scf}\right|n^{\prime},\textbf{k}+\textbf{q}\right\rangle \nonumber \\
\times \sum_{\textbf{q}\lambda}\left(\frac{1}{2M_{s}\omega_{\textbf{q}\lambda}}\right)^{\frac{1}{2}}e^{-iq\cdot\tau_{s}}\epsilon^{\ast}\left(\frac{\textbf{q}\lambda}{s}\right)
\end{eqnarray}
where, $g_{n^{\prime}n\textbf{k}}^{\textbf{q}\lambda}$ describes the scattering probability from  $\left|\textbf{n},\textbf{k}\right\rangle$ to $\left|\textbf{n}^{\prime},\textbf{k}+\textbf{q}\right\rangle$ as a result of emission or absorption of a phonon with momentum \textbf{q}, frequency $\omega$ in branch $\lambda$. $\phi_{scf}$ is the self-consistent potential obtained by calculating the charge density from DFT. $\alpha$ are the atomic displacements and $\tau_{s}$ is the location of mass $M$ of the $s^{th}$ atomic species in the unit cell with the polarization vectors $\epsilon^{*}\left(\frac{\textbf{q}\lambda}{s}\right)$. DFPT is then used to solve Eq. (A.1) taking 200 random \textbf{q} points in the irreducible BZ. The corresponding energy shift of the state $\left|\textbf{n},\textbf{k}\right\rangle$ can now be obtained from the MBPT calculations. The single particle interacting Green's propagator in this case is $G_{n\textbf{k}}\left(\omega\right)=\left[\omega-\epsilon_{n\textbf{k}}-\sideset{}{_{n\textbf{k}}^{Fan}}\sum\left(\omega\right)-\sideset{}{_{n\textbf{k}}^{DW}}\sum\right]^{-1}$ in which $\epsilon_{n\textbf{k}}$ is the bare energy. $\sideset{}{_{n\textbf{k}}^{Fan}}\sum\left(\omega\right)$ and $\sideset{}{_{n\textbf{k}}^{DW}}\sum$ are the Fan and the Debye-Waller self-energies respectively, that composed of all possible type of scatterings. The former is frequency dependent and can be written as \cite{Fan1950}
\begin{eqnarray}
\sideset{}{_{n\textbf{k}}^{Fan}}\sum\left(\omega\right)&=&\sum_{n^{\prime}\textbf{q}\lambda}\left|g_{n^{\prime}n\textbf{k}}^{\textbf{q}\lambda}\right|^{2}\left[\frac{N\left(\omega_{\textbf{q}\lambda}\right)+1-f_{n^{\prime}\textbf{k}-\textbf{q}}}{\omega-\epsilon_{n^{\prime}\textbf{k}-\textbf{q}}-\omega_{\textbf{q}\lambda}-i0^{+}} \right. \nonumber \\ &+& \left. \frac{N\left(\omega_{\textbf{q}\lambda}\right)+f_{n^{\prime}\textbf{k}-\textbf{q}}}{\omega-\epsilon_{n^{\prime}\textbf{k}-\textbf{q}}+\omega_{\textbf{q}\lambda}-i0^{+}}\right]
\end{eqnarray}
in which $N$ and $f$ are the Bose and Fermi functions respectively. The later is the frequency independent and can be expressed as \cite{Cannuccia2013, Cannuccia2011}
\begin{equation}
\sideset{}{_{n\textbf{k}}^{DW}}\sum=-\sum_{\textbf{q}\lambda}\sum_{n^{\prime}}\frac{\Lambda_{nn^{\prime}\textbf{k}}^{\textbf{q}\lambda,-\textbf{q}\lambda}}{\epsilon_{n\textbf{k}}-\epsilon_{n^{\prime}\textbf{k}}}\left[2N\left(\omega_{\textbf{q}\lambda}\right)+1\right]
\end{equation}
The coefficients $\Lambda_{nn^{\prime}k}^{\textbf{q}\lambda,-\textbf{q}\lambda}$ are the second order couplings $=\frac{1}{2}{\textstyle \sum_{s}\sum_{\alpha,\beta}\frac{\epsilon_{\alpha}^{*}\left(\frac{\textbf{q}\lambda}{s}\right)\epsilon_{\beta}\left(\frac{-\textbf{q}\lambda}{s}\right)}{2M_{s}\omega_{\textbf{q}\lambda}}\left\langle n\textbf{k}+\textbf{q}+\textbf{q}^{\prime}\right.}$ $\left.\left|\nabla_{\alpha s}\nabla_{\beta s}\phi_{scf}\right|n\textbf{k}\right\rangle $.
We note here that in order to calculate the Debye-Waller term, a second-order derivative of the self-consistent potential within the perturbation theory is required. This is extremely computationally costly and is not provided by the DFPT calculation. Thus, in practice one uses the rigid-ion approximation and re-cast the Debye-Waller in terms of a product of Fan-like terms \cite{Ponc2014, Ponc2015}. Such modification needs Sternheimer linear solution \cite{Sternheimer1954} to avoid summation over empty electronic states. The MBPT Yambo code does not have such Sternheimer implementation. As a result the zero-point renormalization (ZPR) will converge very slowly with the number of bands. Therefore, in spirit of this we provide a ZPR convergence with the number of electronic bands in Fig. S6 of \cite{Supplemental}.\\
Using $G_{n\textbf{k}}\left(\omega\right)$ and Eqs. (A.2) and (A.3), it is now possible to write the energy shift $\Delta E_{n\textbf{k}}$ of the state $\left|\textbf{n},\textbf{k}\right\rangle$ as \cite{Marini2008}
\begin{equation}
\Delta E_{n\textbf{k}}-\epsilon_{n\textbf{k}}\approx Z_{n\textbf{k}}\Re\left[\sideset{}{_{n\textbf{k}}^{Fan}}\sum\left(\omega\right)+\sideset{}{_{n\textbf{k}}^{DW}}\sum\right]
\end{equation} 
in which $Z_{n\textbf{k}}=\left[1-\left.\frac{\partial}{\partial\omega}\Re\sum_{n\textbf{k}}^{Fan}\left(\omega\right)\right|_{\omega=\epsilon_{n\textbf{k}}}\right]^{-1}$ is the QP renormalized weight factor ($0 < Z_{n\textbf{k}} \leq 1$) in this case. $Z_{n\textbf{k}}\rightarrow1$ for $\frac{\partial}{\partial\omega}\Re\sum_{n\textbf{k}}^{Fan}\rightarrow0$ is known as the static or the on-the-mass-shell approximation \cite{Marini2008}. Once $G_{n\textbf{k}}\left(\omega\right)$ is known, the spectral function can then be expressed as
\begin{equation}
A_{n,\textbf{k}}\left(\omega,T\right)=\frac{1}{\pi}\frac{\left|\Im\sum^{ep}\left(\omega\right)\right|}{\left[\omega-\epsilon_{nk}-\Re\sum^{ep}\left(\omega\right)\right]^{2}+\left[\Im\sum^{ep}\left(\omega\right)\right]^{2}}
\end{equation}
in which $\sum^{ep}\left(\omega\right)=\sideset{}{_{n\textbf{k}}^{Fan}}\sum\left(\omega\right)+\sideset{}{_{n\textbf{k}}^{DW}}\sum$. The shifts can be used to determine the Eliashberg function at each state $\left|\textbf{n},\textbf{k}\right\rangle$ as \cite{Mahan2014}
\begin{equation}
g_{n\textbf{k}}^{2}F\left(\omega\right)=\sum_{q\lambda}\frac{\partial E_{n\textbf{k}}}{\partial N\left(\omega_{q\lambda}\right)}\delta\left(\omega-\omega_{q\lambda}\right)
\end{equation}
Equation (A.6) can be computed at any state to get the required difference as shown in Fig. (2-c).\\
Under linear response theory, electron-electron interaction is both correlated and exchanged. The correlation part is long range and dynamic. The corresponding matrix element of the self-energy in the plane wave basis set is diagonal and can be expressed as \cite{Rohlfing2000, Mahan2014}
\begin{widetext}
\begin{equation}
\left\langle n\textbf{k}\left|\sum^{ee}\left(\omega\right)\right|n\textbf{k}\right\rangle =i\sum_{m}\int_{BZ}\frac{d\textbf{q}}{\left(2\pi\right)^{3}}\sum_{\textbf{GG}^{\prime}}\frac{4\pi}{\left|\textbf{q}+\textbf{G}\right|^{2}}\rho_{nm}\left(\textbf{k},\textbf{q},\textbf{G}\right)\rho_{nm}^{\ast}\left(\textbf{k},\textbf{q},\textbf{G}\right)\int d\omega^{\prime}G_{m\textbf{k}-\textbf{q}}^{0}\left(\omega-\omega^{\prime}\right)\varepsilon_{\textbf{GG}^{\prime}}^{-1}\left(\textbf{q},\omega^{\prime}\right)
\end{equation}
\end{widetext}
Note that, computationally this diagonal matrix evaluation depends on the convergence of $m$ the number of electronic bands, BZ integral and the \textbf{G} the G-vectors in the Coulomb potential in the Fourier transformed plane. $\varepsilon_{\textbf{GG}^{\prime}}^{-1}$ is the microscopic dynamic dielectric function that is evaluated numerically from the generalized plasmon-pole model \cite{Godby1989} and $G^{0}_{m,\textbf{k}-\textbf{q}}$ is the non-interacting Green's propagator. The other portion of the self-energy is exchange or static one (the Hartree-Fock self-energy) and is also diagonal. The matrix elements are $\left\langle nk\left|\sum^{x}\right|n\textbf{k}\right\rangle =-\sum_{m}\int_{BZ}\frac{d\textbf{q}}{\left(2\pi\right)^{3}}\sum_{\textbf{G}}\frac{4\pi}{\left|\textbf{q}+\textbf{G}\right|^{2}}\left|\rho_{nm}\left(\textbf{k},\textbf{q},\textbf{G}\right)^{2}\right|f_{m,\textbf{k}-\textbf{q}}$. The total self-energy is thus $\left\langle n\textbf{k}\left|\sum^{ee}\right|n\textbf{k}\right\rangle +\left\langle n\textbf{k}\left|\sum^{x}\right|n\textbf{k}\right\rangle$ leading to the QP energy as 
\begin{widetext}
\begin{equation}
E_{n\textbf{k}}=\epsilon_{n\textbf{k}}+Z_{n\textbf{k}}\Re\left[\left\langle \psi_{n\textbf{k}}\left|\left\langle n\textbf{k}\left|\sum^{ee}\right|n\textbf{k}\right\rangle +\left\langle n\textbf{k}\left|\sum^{x}\right|n\textbf{k}\right\rangle -V_{xc}\right|\psi_{n\textbf{k}}\right\rangle \right]
\end{equation}
\end{widetext}
in which $V_{xc}$ is the exchange-correlation functional at the LDA level. The QP lifetimes are the reciprocal of the imaginary part of $\left\langle n\textbf{k}\left|\sum^{ee}\right|n\textbf{k}\right\rangle +\left\langle n\textbf{k}\left|\sum^{x}\right|n\textbf{k}\right\rangle$. $Z_{n\textbf{k}}<$1 \cite{Mahan2014} is the QP renormalized weight factors in this case. $E_{n\textbf{k}}$ from Eq. (A.8) and $\Delta E_{n\textbf{k}}$ from Eq. (A.4) when added gives the valence band maxima and conduction band minima separation due to both G$_{0}$W$_{0}$ and lattice vibrations.\\
Excitons are the strongly correlated two-particle electron-hole system and are attracted by dynamic long-range Coulombic force. The motion is governed by the two-particle Schr{\"o}dinger equation, also known as BSE. The hamiltonian is hermitian in the absence of lattice vibration and can be written as \cite{Marini2008}
\begin{equation}
H_{ee^{\prime},hh^{\prime}}=\left(E_{e}-E_{h}\right)\delta_{eh,e^{\prime}h^{\prime}}+\left(f_{e}-f_{h}\right)K_{ee^{\prime},hh^{\prime}}
\end{equation}
The BSE kernel is $K_{ee^{\prime},hh^{\prime}}$ which is a sum of a bare positive Coulomb exchange term (repulsive) and a direct negative screened electron-hole interaction (attractive). The repulsive part is due to the Hartree potential variation and causes the spin singlet or triplet splitting while the attractive part is long range and causes excitons to form. The exciton wave-functions are the eigen-states $\left|\varphi\right\rangle$ in the electron-hole pair basis space ($e$-$h$). $E_{e}$ and $E_{h}$ are electron-hole QP energies which adds to the broadening part ($\Delta E_{e}\left(T\right)$ and $\Delta E_{h}\left(T\right)$) when lattice vibrations are included. This addition makes the hermitian hamiltonian to a non-hermitian. The energy eigenvalues can then be written as \cite{Marini2008}
\begin{eqnarray}
E_{\varphi}\left(T\right)=\left\langle \varphi\left(T\right)\left|H\right|\varphi\left(T\right)\right\rangle +\sum_{e,h}\left|A_{e,h}^{\varphi}\left(T\right)\right|^{2} \nonumber \\ 
\times\left[\Delta E_{e}\left(T\right)-\Delta E_{h}\left(T\right)\right]
\end{eqnarray}
where $\left|\varphi_{FA}\left(T\right)\right\rangle $ are now temperature dependent with $\left|\varphi\left(T\right)\right\rangle =\sum_{e,h}A_{e,h}^{\varphi}\left(T\right)\left|e,h\right\rangle $. As usual, the coefficients $A_{e,h}^{\varphi}\left(T\right)=\left\langle e,h\left|\varphi\left(T\right)\right.\right\rangle $
are the inner products. The real and imaginary part of the exciton energies are \cite{Marini2008}
\[
\Re\left[\Delta E_{\varphi}\left(T\right)\right]=\left\langle \varphi\left(T\right)\left|H\right|\varphi\left(T\right)\right\rangle -\left\langle \varphi\left|H\right|\varphi\right\rangle 
\]
\begin{equation}
+\int d\omega\Im\left[g^{2}F_{\varphi}\left(T\right)\right]\left[N\left(\omega,T\right)+\frac{1}{2}\right]
\end{equation} 
and 
\begin{equation}
\Im\left[E_{\varphi}\left(T\right)\right]=\int d\omega\Im\left[g^{2}F_{\varphi}\left(T\right)\right]\left[N\left(\omega,T\right)+\frac{1}{2}\right]
\end{equation}  
The exciton-phonon coupling function is defined by the difference $g^{2}F_{\varphi}\left(T\right)=\sum_{e,h}\left|A_{eh}^{\varphi}\left(T\right)\right|^{2}\left[g^{2}F_{e}\left(\omega\right)-g^{2}F_{h}\left(\omega\right)\right]$ with $\Delta E_{\varphi}\left(T\right)=E_{\varphi}\left(T\right)-E_{\varphi}$. We note here that as T$\rightarrow$0 Eq. (A.11) remains finite and is in accordance with the Heisenberg's uncertainty principle. This produces a finite lifetime, which would be infinite in frozen atom condition. The macroscopic dielectric function in the long wavelength (\textbf{q}$\rightarrow$0) is now a temperature dependent function with no fitting or broadening parameter
\begin{equation}
\varepsilon_{M}\left(\omega,T\right)=-\frac{8\pi}{\Omega}\sum_{\varphi}\left|O_{\varphi}\left(T\right)\right|^{2}\Im\left(\omega-E_{\varphi}\left(T\right)\right)^{-1}
\end{equation}
$\Im\varepsilon_{M}\left(\omega,T\right)$ is the temperature dependent absorption spectra and the quantity  $O_{\varphi}\left(T\right)=\left\langle n\textbf{k}\left|exp\left(i\boldsymbol\kappa\cdot\mathbf{r}\right)\left(\left[\left.\left|\varphi\left(T\right)\right.\right\rangle -\left.\left|\varphi\right.\right\rangle \right]\right)\right.\right.$ is the corresponding oscillator strength, $\boldsymbol\kappa$ is the polarization vector direction and $\Omega$ is the unit-cell volume. The reader is encouraged to the references \cite{Rohlfing2000, Fetter2013, Marini2008, Cannuccia2013, Elena2011, Mahan2014} for a more detailed theory.
\nocite{*}

\bibliography{apssamp}
\end{document}



\title{Giant exciton-phonon coupling and zero-point renormalization in hexagonal monolayer boron nitride}
\author{Himani Mishra${}^\ddagger$}
 \author{Sitangshu Bhattacharya${}^\ddagger$}
 \email{Corresponding Author's Email: sitangshu@iiita.ac.in}
\affiliation{${}^\ddagger$Nanoscale Electro-Thermal Laboratory, Department of Electronics and Communication Engineering, Indian Institute of Information Technology-Allahabad, Uttar Pradesh 211015, India}
\pacs{78.20.Ci, 71.35.Cc, 71.35.-y, 63.20.Ls, 31.15.Md, 11.10.St, 71.35.Aa, 63.20.kk}
\keywords{Monolayer hBN; electron-phonon coupling; polaronic widths; Bethe-Salpeter equation; excitons; spectral functions; non-radiative linewidths}

\maketitle
\onecolumngrid

\newpage \section{Kinetic energy cut-off convergence}
\begin{figure*}
  \includegraphics[width=0.95\columnwidth]{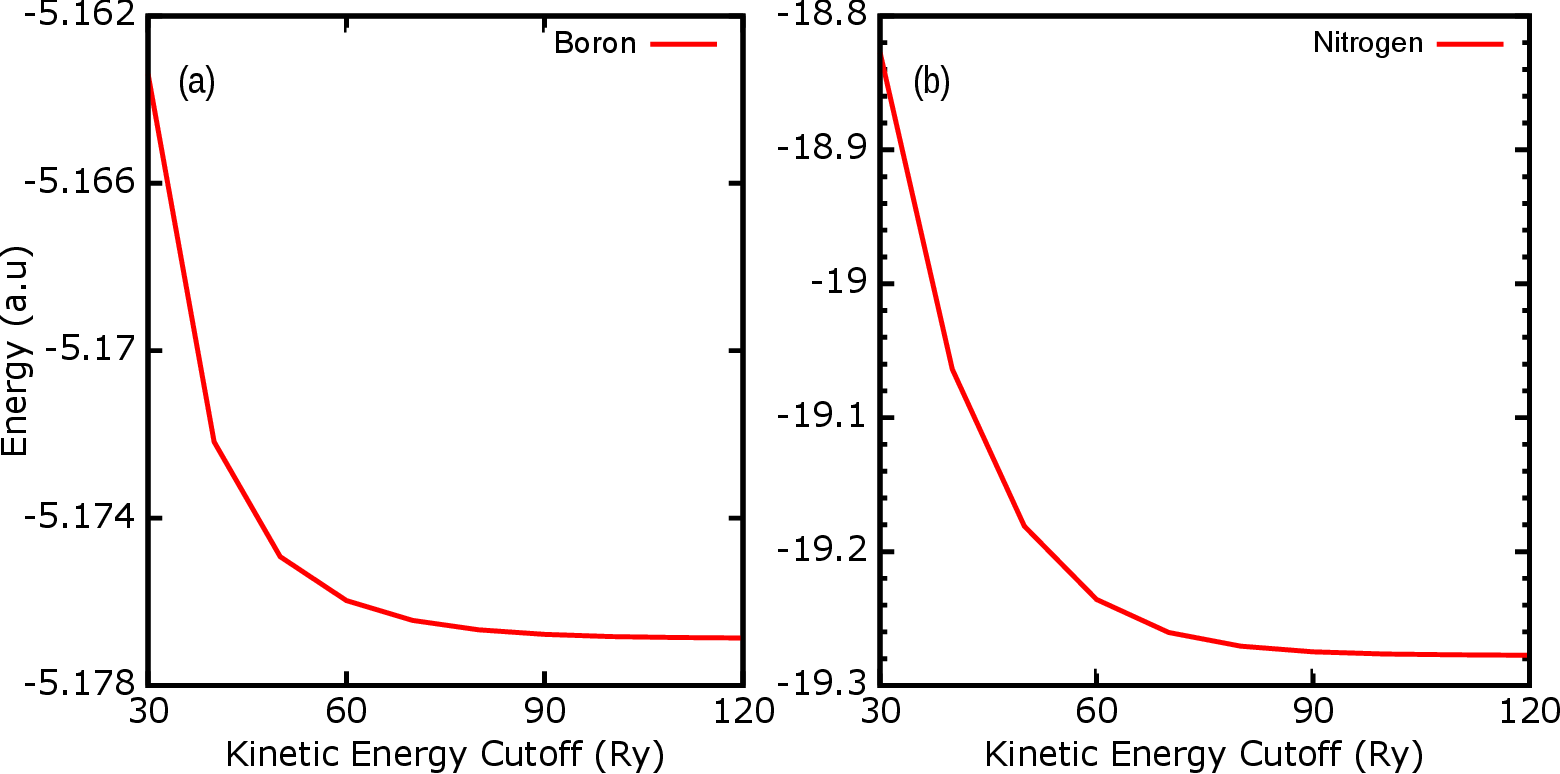}
  \caption{Total energy minimization as function of kinetic energy cut-off using a norm-conserving pseudo-potential of (a) boron and (b) nitrogen. We see that the energy settles from 90 Ry in both cases. We therefore choose 120 Ry in all our computation.}
  \label{fgr:example}
\end{figure*}

\newpage 
\section{Convergence with respect to k-point sampling for determination of frozen-atom ground state band structure along high symmetries of BZ.}
\begin{figure*}
  \includegraphics[width=0.95\columnwidth]{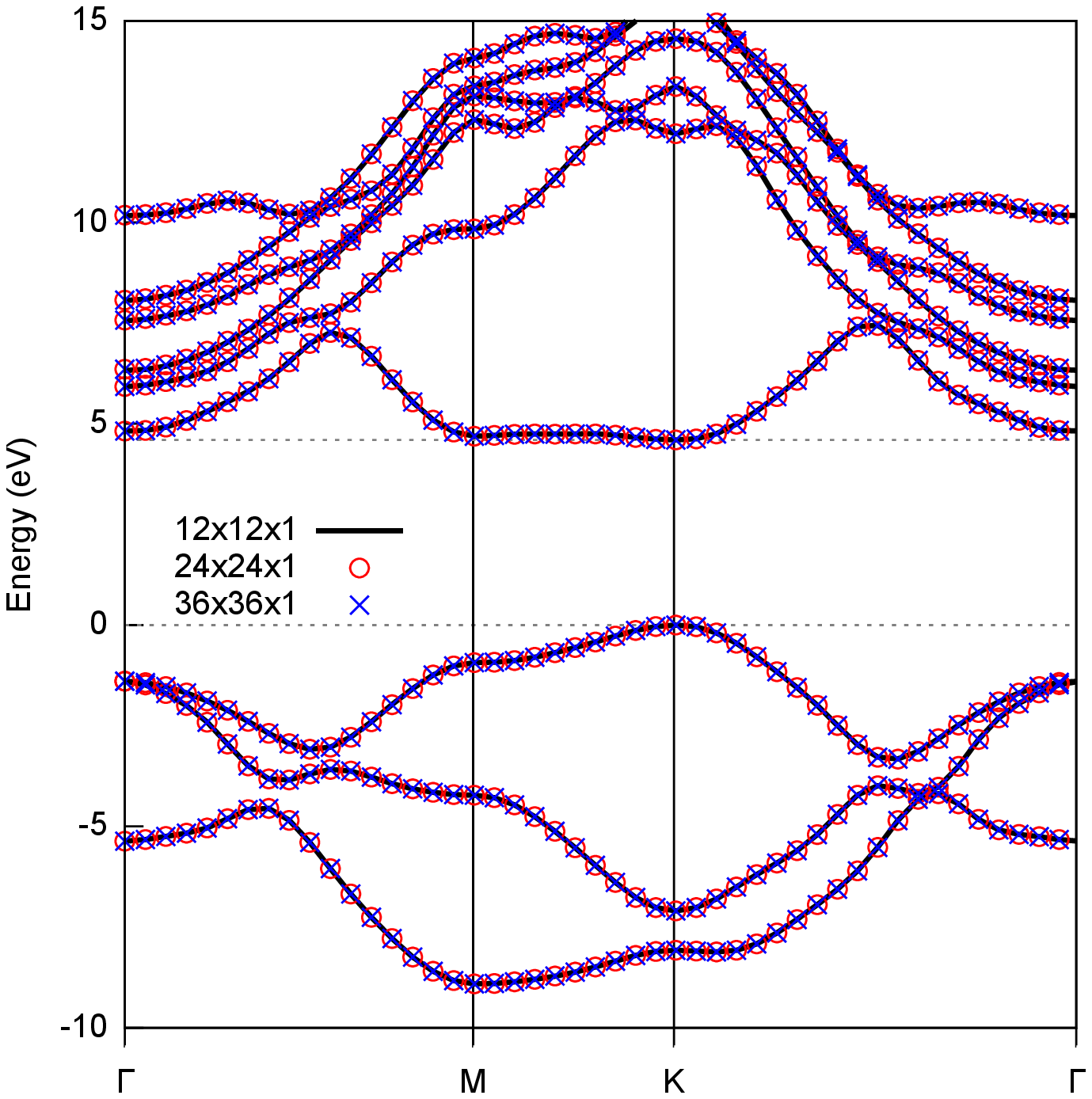}
  \caption{Frozen atom ground state band-structure of ML hBN showing convergence with respect to k-point sampling. We see that the results are well-converged from 12$\times$12$\times$1 to a dense 36$\times$36$\times$1. In all the sampling cases, the energy band-gap is direct at \textbf{K}. We thus choose a 12$\times$12$\times$1 for both DFT and G$_{0}$W$_{0}$ calculations with the cut-off of 120 Ry, as demonstrated in Fig. S1.}
  \label{fgr:example}
\end{figure*}

\newpage 
\section{G$_{0}$W$_{0}$ QP energy stretching as function of bare energy}
\begin{figure*}
  \includegraphics[width=0.8\columnwidth]{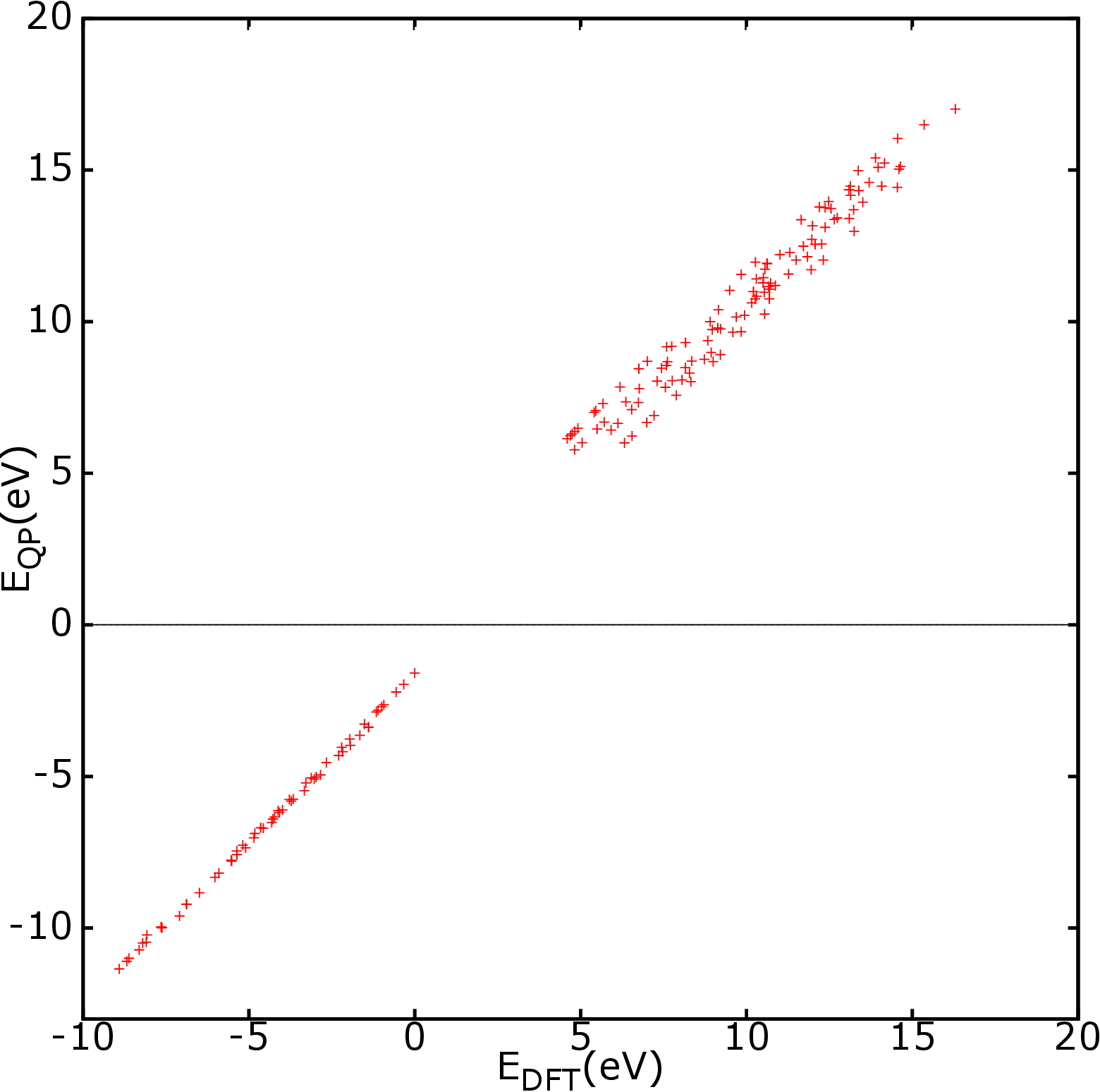}
  \caption{G$_{0}$W$_{0}$ QP energies are given as a function of DFT-LDA energies. The top of DFT valence band is located at 0 level. In order to determine the fit for the valence and conduction bands, we apply data regression technique and evaluate their spreading in the case of G$_{0}$W$_{0}$ in comparison to DFT values. To account for the rigid shift we give a scissor of 3.129 eV and spreading of valence and conduction bands are about 1.26 eV and 1.07 eV respectively.}
  \label{fgr:example}
\end{figure*}

\newpage 
\section{Convergence with respect to k-point sampling and number of bands for the determination of optical absorption spectra.}
\begin{figure*}
  \includegraphics[width=0.85\columnwidth]{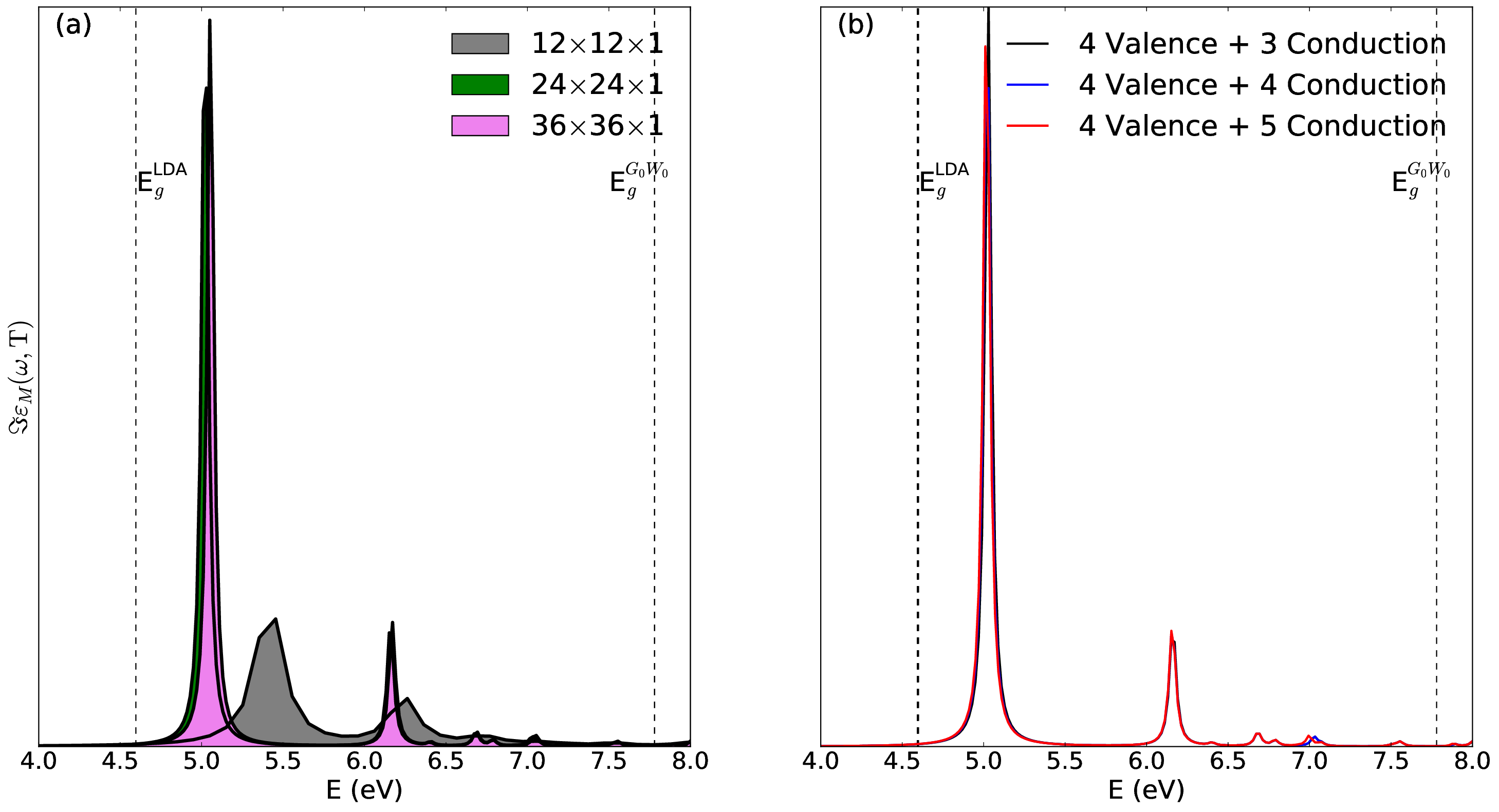}
  \caption{(a) Absorption spectra using frozen atom G$_{0}$W$_{0}$ results done on 12$\times$12$\times$1 (gray), 24$\times$24$\times$1 (green) to a dense 36$\times$36$\times$1 (magenta) k-point sampling. We observe that the spectrum is grossly under-estimated for the 12$\times$12$\times$1 k-point sampling. A reasonably good convergence is achieved with 24$\times$24$\times$1 sampling with respect to 36$\times$36$\times$1. This is in excellent agreement with the optical spectra convergence done with 24$\times$24$\times$1 \cite{Rasmussen2016, Galvani2016, Paleari2018}. (b) Convergence of absorption spectra with respect to the number of bands involved in transition. We use 24$\times$24$\times$1 sampling data of Fig. S4 (a). In our calculation, we used 4 highest valence bands and 3, 4 and 5 lowest conduction bands respectively. An excellent convergence is achieved for all the cases, signifying all the valence bands and lowest 4 conduction bands are sufficient to get converged spectra.}  
  \label{fgr:example}
\end{figure*}

\newpage 
\section{$p_{z}$-Orbital weightage of boron and nitrogen in the band structure along high symmetries of BZ.}
\begin{figure*}
  \includegraphics[width=0.95\columnwidth]{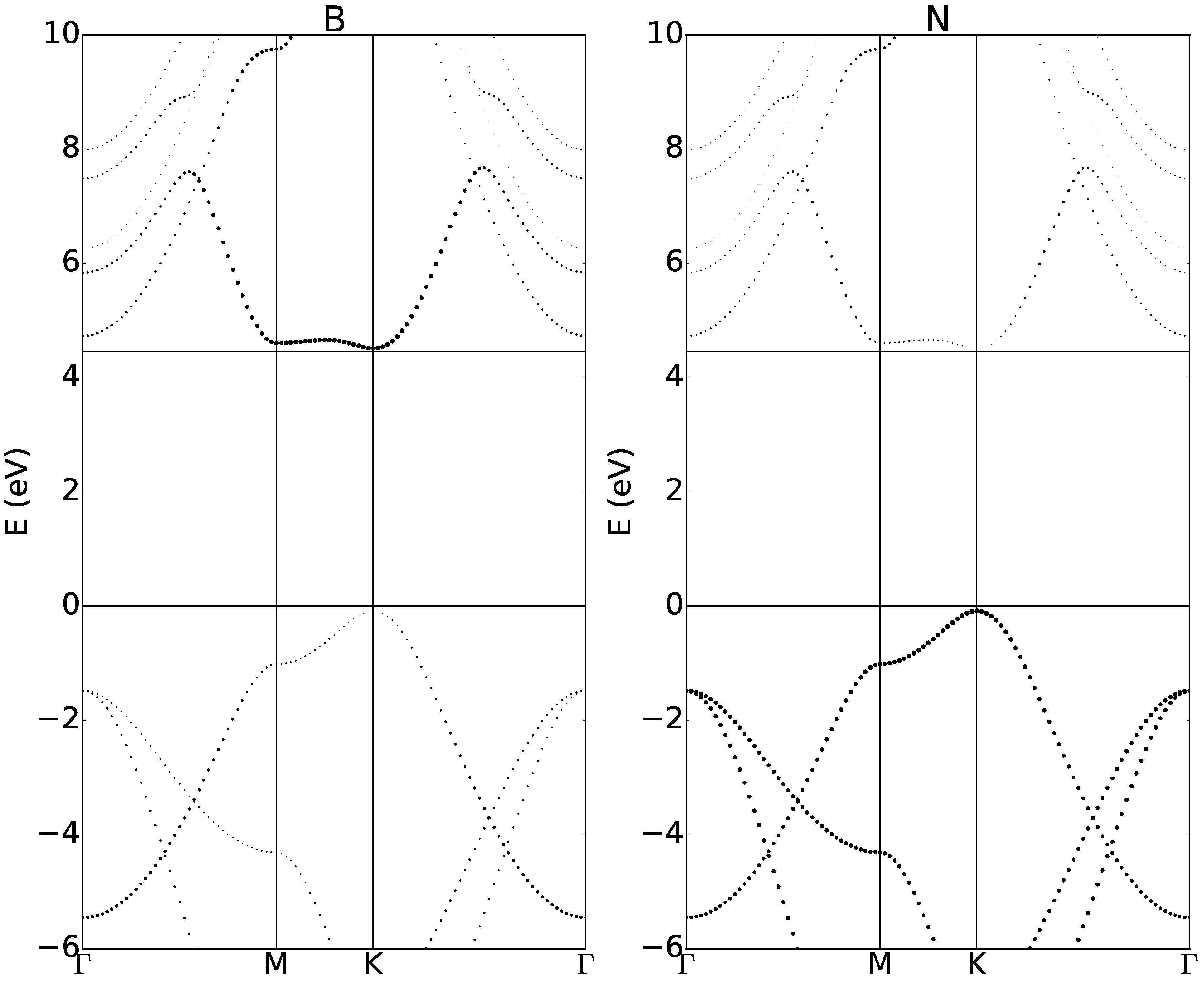}
  \caption{The $p_{z}$-orbital weightage of (a) boron atoms and (b) nitrogen atoms on the frozen atom ground state band-structure. The size of the points are the corresponding weights.}
  \label{fgr:example}
\end{figure*}

\newpage 
\section{Spectral function convergence with respect to electronic bands.}
\begin{figure*}
  \includegraphics[width=0.85\columnwidth]{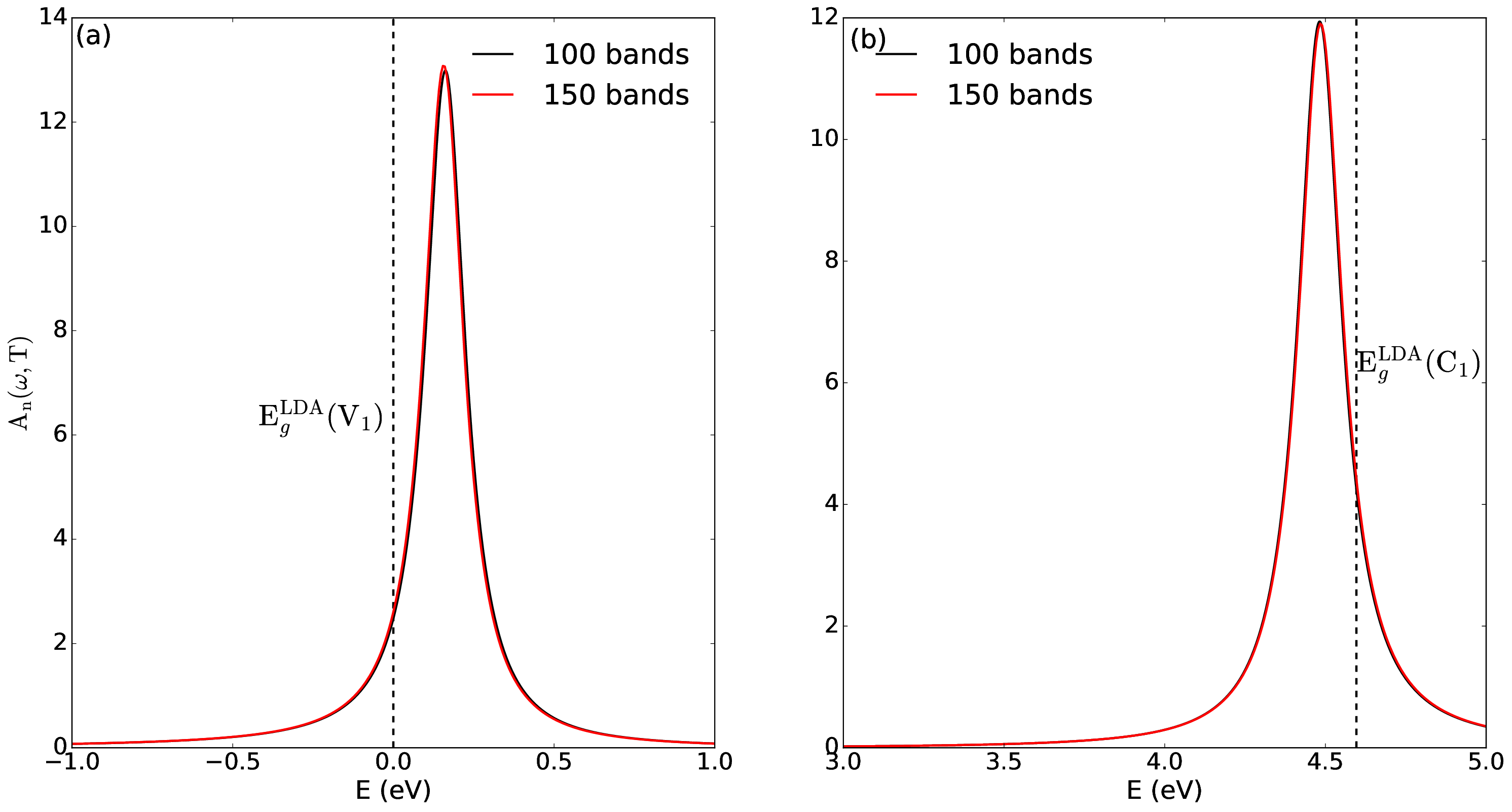}
  \caption{Spectral function convergence with respect to the electronic bands at T = 0 K. (a) top valence band and (b) bottom conduction band at K in the BZ (Please see Fig. 1(a) in the main text for the notations).}
  \label{fgr:example}
\end{figure*}

\newpage 
\section{Exciton wave-function of single lowest lying bright state}
\begin{figure*}
  \includegraphics[width=0.85\columnwidth]{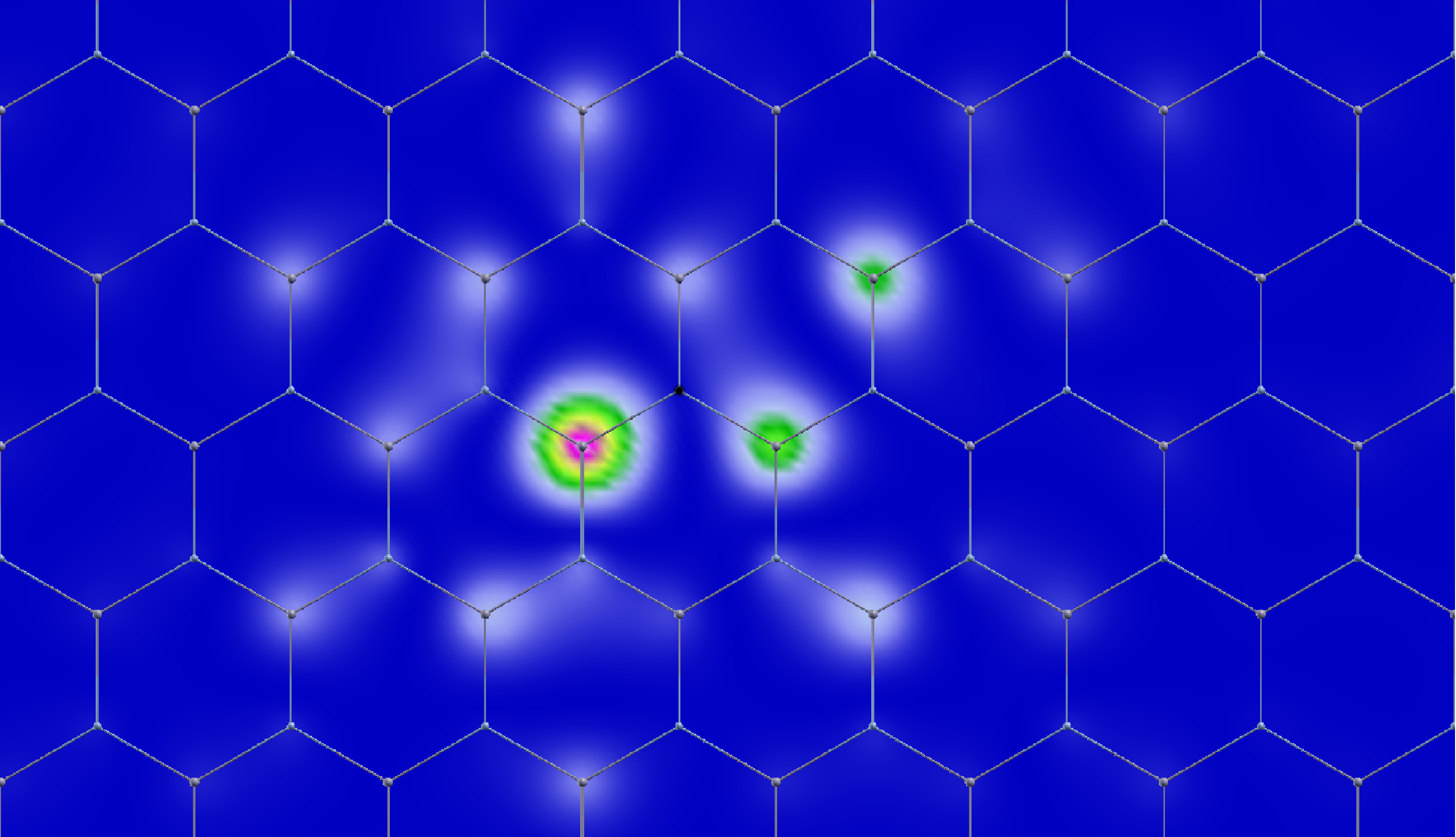}
  \caption{Excitonic wave-function plot at 300 K over the xy plane of the ML h-BN when only bright exciton is used. The hole is fixed in space (black dot) and is placed on the top of the N atom at a distance of 1 $\mathring{\mathrm{A}}$. Here we can understand the effect of degeneracy in the main excitonic peak as only one bright exciton state is contributing to the wave-function. The three fold rotation symmetry \cite{Arnaud2006, Wirtz2008} can be recovered by adding the other degenerate excitonic wave-function (See the main text).}
  \label{fgr:example}
\end{figure*}

\newpage 
\section{Band-structure of monolayer hBN with spinor component in the wave function}
\begin{figure*}
  \includegraphics[width=0.65\columnwidth]{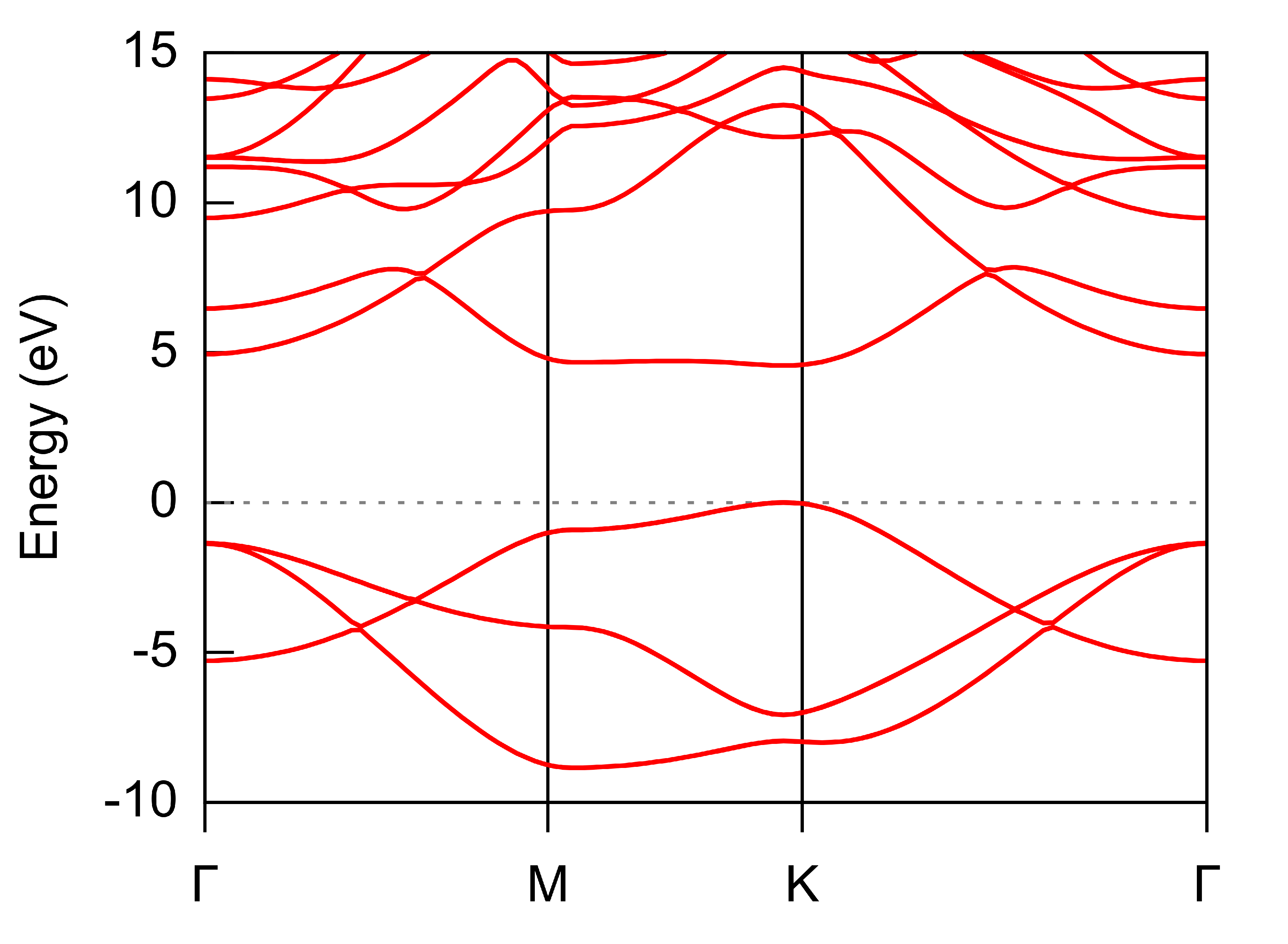}
  \caption{A spin-dependent calculation of energy band-structure of hBN using a fully relativistic and norm-conserving pseudo-potential at the LDA level. No spin-splitting in valence or conduction bands were found. This justifies the use of negligible spinor component in our calculations.}
  \label{fgr:example}
\end{figure*}
%
\bibliography{apssamp_supple}